%% file: HIN-21-004_temp.tex
\pdfoutput=1
\documentclass[11pt,twoside,a4paper,cmspaper,final,collab]{cms-tdr}
\def\svnVersion{5c847f0}\def\svnDate{2024/01/29}\def\cmsCernNoTag{CERN-EP-2023-085}\def\cmsCernDate{\today}\def\cmsMessage{Published in the Journal of High Energy Physics as \href{http://dx.doi.org/10.1007/JHEP01(2024)128}{\doi{10.1007/JHEP01(2024)128}.}}
\begin{document}\cmsNoteHeader{HIN-21-004}

\newlength\cmsTabSkip\setlength{\cmsTabSkip}{1ex}
\newcommand{\PbPb}{\ensuremath{\text{PbPb}}\xspace}
\newcommand{\LcDratio}{\ensuremath{\PcgLp/ \PDz}\xspace}
\newcommand{\Lcdecay}{\ensuremath{\PcgLp\to\Pp\PKm\Pgpp}\xspace}
\newcommand{\RAA}{\ensuremath{R_{\mathrm{AA}}}\xspace}
\newcommand{\taa}{\ensuremath{T_{\mathrm{AA}}}\xspace}
\newcommand{\absy}{\ensuremath{\abs{y}}\xspace}
\newcommand{\Ncoll}{\ensuremath{N_\text{coll}}\xspace}
\newcommand{\Npart}{\ensuremath{N_\text{part}}\xspace}
\newcommand{\abseta}{\ensuremath{ \abs{\eta}}\xspace}
\newcommand{\Raap}{\ensuremath{R_{\mathrm{AA}}^{\text{prompt}}}}
\newcommand{\Raanon}{\ensuremath{R_{\mathrm{AA}}^{\text{nonprompt}}}}
\newcommand{\Raaratio}{\ensuremath{\Raanon/ \Raap}\xspace}
\newcommand{\Vz}{\ensuremath{V_\mathrm{z}}\xspace}

\cmsNoteHeader{HIN-21-004}
\title{Study of charm hadronization with prompt \texorpdfstring{\PcgLp}{Lambda(c)} baryons in proton-proton and lead-lead collisions at \texorpdfstring{\sqrtsNN=5.02\TeV}{sqrt(sNN) = 5.02 TeV}}

\date{\today}

\abstract{The production of prompt \PcgLp baryons is measured via the exclusive decay channel \Lcdecay at a center-of-mass energy per nucleon pair of 5.02\TeV, using proton-proton ($\Pp\Pp$) and lead-lead (\PbPb) collision data collected by the CMS experiment at the CERN LHC. The $\Pp\Pp$ and \PbPb data were obtained in 2017 and 2018 with integrated luminosities of 252 and 0.607\nbinv, respectively. The measurements are performed within the \PcgLp rapidity interval $\absy<1$ with transverse momentum (\pt) ranges of 3--30 and 6--40\GeVc for $\Pp\Pp$ and \PbPb collisions, respectively. Compared to the yields in $\Pp\Pp$ collisions scaled by the expected number of nucleon-nucleon interactions, the observed yields of \PcgLp with $\pt>10\GeVc$ are strongly suppressed in \PbPb collisions. The level of suppression depends significantly on the collision centrality. The \LcDratio production ratio is similar in \PbPb and $\Pp\Pp$ collisions at $\pt>10\GeVc$, suggesting that the coalescence process does not play a dominant role in prompt \PcgLp baryon production at higher \pt.}

\hypersetup{pdfauthor={CMS Collaboration},pdftitle={Study of charm hadronization with prompt \texorpdfstring{\PcgLp}{Lambda(c)} baryons in proton-proton and lead-lead collisions at \texorpdfstring{\sqrtsNN=5.02\TeV}{sqrt(sNN) = 5.02 TeV}},pdfsubject={CMS},pdfkeywords={CMS, physics, Lambda(c), nuclear modification factor, heavy flavor, prompt}}

\maketitle

\section{Introduction}
In relativistic heavy ion collisions, hadrons containing a heavy quark are versatile probes for understanding parton energy loss and thermalization properties associated with the quark-gluon plasma (QGP) formed in these collisions~\cite{QGP, Andronic:2015wma,Prino:2016cni,Dong:2019unq,Dong:2019byy,Busza:2018rrf, Apolin_rio_2022}. Because of their large rest masses, heavy quarks are primarily produced in the initial hard scatterings and interact with the medium by different energy loss mechanisms than light-flavor quarks~\cite{energyloss}. A thorough understanding of both the in-medium interactions and the subsequent hadronization processes is needed to correctly interpret the experimental data. In proton-proton ($\Pp\Pp$) collisions, hadronization is expected to occur in the form of a fragmentation process~\cite{ANDERSSON198331}. While the fragmentation process can still occur during heavy ion collisions, in these collisions hadrons can also form by parton coalescence inside or at the boundary of the QGP medium~\cite{Greco_introduction,Greco_follow}. The effect of coalescence is expected to be most prominent for low and intermediate transverse momentum (\pt) hadrons, for which the density is highest for the precursor partons. The coalescence probability decreases with increasing hadron \pt, leading to the dominance of the fragmentation process of hadronization at high \pt.

In the presence of a QGP, the parton coalescence contribution for baryon production is expected to be higher than that for meson production at a fixed hadron \pt value. This is a consequence of the hadron \pt value reflecting the sum of the constituent quark \pt values and of having a decreasing quark density with increasing quark \pt. In particular, a large enhancement in the \LcDratio yield ratio in heavy ion relative to $\Pp\Pp$ collisions is predicted by models including the coalescence of charm and light-flavor quarks in \PcgLp baryon production~\cite{Oh09,Lee,ghosh,Greco_publication}, with the enhancement expected to have a strong \pt dependence.  

The production of \PcgLp baryons for a variety of collision systems has been measured by RHIC and LHC experiments. At the LHC, \PcgLp \pt-differential cross section measurements have been made for $\Pp\Pp$ collisions at midrapidity ($y\sim0$) by ALICE~\cite{ALICE_pp, ALICE_pp2} and CMS~\cite{HIN18009}, and at forward rapidity by LHCb~\cite{LHCb_pp}. These measurements are compared to theoretical predictions from the next-to-leading order general-mass variable-flavor-number scheme (GM-VFNS)~\cite{GM-VFNS,gmvfns1} and to $\PYTHIA 8$ with color reconnection~\cite{CRmodes}. Within the rapidity range 2.0--4.5, these predictions are consistent with the LHCb results. The GM-VFNS calculations, however, underestimate ALICE and CMS results for $\absy<0.5$ and 1.0, respectively. The $\PYTHIA 8$ calculations with mode 2 of the color reconnection mechanism (CR2)~\cite{CRmodes} activated are in general agreement with the experimental results~\cite{ALICE_pp2,HIN18009}. ALICE and LHCb also report \LcDratio production ratios for $\Pp\Pp$ and proton-lead ($\Pp$Pb) collisions. The ALICE ratios from $\Pp\Pp$ and $\Pp$Pb collisions~\cite{ALICE_pp} are above the corresponding LHCb values~\cite{LHCb_result}, indicating a rapidity dependence for baryon and meson hadronization.

At RHIC, the STAR Collaboration measured \PcgLp production at midrapidity ($\absy < 1$) in gold-gold collisions at a center-of-mass energy per nucleon pair ($\sqrtsNN$) of 200\GeV~\cite{STAR_Lc}. In heavy ion collisions, measurements are often done in centrality classes that are given in percentage ranges of the total inelastic hadronic cross section, with the 0--20\% centrality bin corresponding to the 20\% of collisions having the largest overlap of the two nuclei. Within the 0--20\% centrality range, the \LcDratio ratio for $3<\pt<6\GeVc$ is found to be significantly higher than predicted by the $\PYTHIA 8$ event generation model~\cite{PYTHIA8} for $\Pp\Pp$ collisions. Also, an ALICE measurement of the \LcDratio ratio for $\pt<10\GeVc$ at midrapidity in lead-lead (\PbPb) collisions at $5.02\TeV$~\cite{ALICE_PbPb, ALICE_PbPb2} finds larger values than those in $\Pp\Pp$ and $\Pp$Pb collisions. In contrast, the CMS results in the higher-\pt range 10--20\GeVc are consistent with the $\Pp\Pp$ results~\cite{HIN18009}. The \PcgLp nuclear modification factor (\RAA) is the ratio of \PcgLp baryon yields in \PbPb to the $\Pp\Pp$ results scaled by the number of nucleon-nucleon (NN) binary collisions. The ALICE and CMS measurements of the \RAA at $5.02\TeV$ find this ratio to be less than unity, indicating a medium-induced charm quark energy loss before hadronization~\cite{ALICE_pp,ALICE_pp2,ALICE_PbPb,HIN18009,ALICE_PbPb2}. More information on the coalescence process in heavy flavor hadron production can be obtained from the recent LHC measurements at $5.02\TeV$ of $\PXP{3872}$ in \PbPb collisions~\cite{CMSX3872}, and \PBpc production in $\Pp\Pp$ and \PbPb collisions~\cite{HIN20004}.

In this paper, we report on the measurements of prompt \PcgLp baryon production in $5.02\TeV$ $\Pp\Pp$ and \PbPb collisions with data collected using the CMS detector in 2017 and 2018, respectively. The integrated luminosities of the $\Pp\Pp$ and \PbPb data are 252 and 0.607\nbinv, respectively~\cite{CMS:2022bjp,CMS-PAS-LUM-19-001,Sirunyan:2759951}. The term ``prompt \PcgLp'' is used to denote \PcgLp baryons that are either directly generated during charm quark hadronization or formed through the strong interaction decay of excited charmed hadron states~\cite{Rapp_ppresult}. This is in contrast to ``nonprompt \PcgLp'', which refers  to \PcgLp baryons that originate from the decay of \cPqb hadrons.

The \PcgLp baryons are reconstructed in the central rapidity region ($\absy<1$) via the hadronic decay channel \Lcdecay. In $\Pp\Pp$ collisions, the \pt spectrum is measured in the \pt range 3--30\GeVc. The \pt spectrum and the \RAA are measured in the integrated centrality class of 0--90\% with $6<\pt<40\GeVc$ and $6<\pt<30\GeVc$, respectively. These are also measured in the centrality bins 0--10, 10--30, 30--50 and 50--90\% within different \pt ranges. The \LcDratio production ratios are obtained for $\Pp\Pp$ and \PbPb collisions with $3<\pt<30\GeVc$ and $6<\pt<40\GeVc$, respectively, using previously reported CMS results for \PDz production~\cite{HIN16001}. The \PbPb \LcDratio ratio is also measured for the centrality class 0--10\% in the \pt range 10--40\GeVc. All references to \PcgLp and \PDz in the text of this paper also include the corresponding charge conjugate states. Tabulated results are provided in the HEPData record for this analysis~\cite{hepdata}.

\section{The CMS apparatus}
The central feature of the CMS apparatus is a superconducting solenoid of 6\unit{m} internal diameter, providing a magnetic field of 3.8\unit{T}. Within the solenoid volume, there is a silicon pixel and strip tracker that measures the trajectories of charged particles in the pseudorapidity range $\abseta<2.5$, a lead-tungstate crystal electromagnetic calorimeter, and a brass and scintillator hadron calorimeter. For charged particles of $1<\pt<10\GeVc$ and $\abseta<1.4$, the typical track resolution is 1.5\% in \pt and 25--90 (45--150)\micron in the transverse (longitudinal) impact parameter~\cite{ppreconstruct}. Two forward hadron (HF) calorimeters use steel as an absorber and quartz fibers as the sensitive material. The two HF calorimeters are located 11.2\unit{m} away from the interaction region, one on either side, and together they extend the calorimeter coverage to $3.0<\abseta<5.2$. Each HF calorimeter consists of 432 readout towers consisting of a geometrically defined group of calorimeter cells. The HF calorimeter cells contain long and short quartz fibers running parallel to the beam that provide information on the shower energy and the relative contribution originating from hadrons versus electrons and photons. In this analysis, the HF information is used for performing an offline event selection and for determining event centrality. The Beam Pickup and Timing for LHC eXperiments (BPTX) provides precise information on the bunch structure and timing of the beam. A detailed description of the CMS experiment can be found in Ref.~\cite{CMS:detector}.

\section{Event selection and simulated event sample}
Events are required to pass selection criteria designed to reject those arising from background processes such as beam-gas interactions and nonhadronic collisions, as described in Ref.~\cite{PbPbreconstruct}. Events must have at least one reconstructed primary interaction vertex formed by two or more tracks~\cite{ppreconstruct} within a distance of 15\cm from the center of the nominal interaction region along the beam axis. In addition, in \PbPb collisions, the shapes of the clusters in the pixel detector must be compatible with those expected from particles produced at the primary vertex location~\cite{ptdistribution}. The \PbPb collision events are also required to have at least two towers in each HF detector with energy deposits of more than 4\GeV per tower. This analysis processes \PbPb minimum-bias (MB) and $\Pp\Pp$ zero-bias (ZB) trigger events. The \PbPb MB trigger requires signals on both sides of the HF calorimeters~\cite{CMS:2016ngn}. The number of interactions per bunch crossing in $\Pp\Pp$ collisions is less than 5, and hence the ZB trigger is used in this analysis. The $\Pp\Pp$ ZB trigger requires a coincidence of the signals from the BPTX detectors located on either side of the interaction point, indicating the presence of bunches from both beams at CMS.

Monte Carlo (MC) simulated event samples for $\Pp\Pp$ and \PbPb collisions are used to optimize the selection criteria, calculate the acceptance times efficiency factors, estimate the systematic uncertainties, and facilitate the separation of prompt and nonprompt \PcgLp baryons. The $\Pp\Pp$ MC samples are generated using $\PYTHIA 8$ 2.1.2~\cite{PYTHIA8}, hereafter referred to as $\PYTHIA 8$, with tune CP5~\cite{Eventune}, which includes both prompt and nonprompt $\PcgLp$ baryon events. For the \PbPb MC samples, each $\PYTHIA 8$ event containing a \PcgLp baryon is embedded into a \PbPb collision event generated with $\HYDJET$ 1.9~\cite{jetquench}, which is tuned to reproduce global event properties such as the charged-hadron \pt spectrum and particle multiplicity. The \Lcdecay decay is simulated using \EVTGEN 1.3.0~\cite{Eventgen} based on four sub-channels: $\PcgLp \to \Pp \PAK^*(892)^0 \to \Pp \PKm \Pgpp$, $\PcgLp \to \PGDppP{1232} \PKm \to \Pp \PKm \Pgpp$, $\PcgLp \to  \PGLP{1520} \Pgpp \to \Pp \PKm \Pgpp$, and \Lcdecay (nonresonant), with no modeling of interference between the sub-channels. Details of the decay channel and its subchannels can be found in Ref.~\cite{Workman:2022ynf}. All particles are propagated through a simulation of the CMS detector response using the \GEANTfour package~\cite{GEANT4}.

\section{Signal reconstruction and extraction}
The signal is reconstructed by selecting three charged-particle tracks within $\abseta<1.2$ in the mass range 2.11--2.45\GeVcc. The net charge of \Lcdecay candidates and their charge conjugates are required to be ${+}1$ and ${-}1$, respectively. All charged tracks must have $\pt>0.5\GeVc$ and 1.0\GeVc for $\Pp\Pp$ and \PbPb collisions, respectively. Signal events will have two same-sign tracks from a proton and a charged pion and an opposite-sign track for the kaon. Without particle identification, the possibilities for either of the same-sign tracks to be a proton or a pion are considered, with the kaon mass assumed for the opposite-sign track. The invariant mass distribution of the reconstructed signal with an incorrect mass assignment is much broader, about 30 times the width of the signal with the correct mass assignment. These misidentified events are indistinguishable from the combinatorial background and are thus treated as part of the background during signal extraction.

The raw signal yields are obtained in each \pt and centrality interval using unbinned maximum likelihood fits. The fit function consists of several components. To account for the variation of track momentum resolution as a function of rapidity, a double Gaussian function is used to model the signal shape in $\Pp\Pp$ collisions, with a common mean value for the two Gaussians, but with different widths. The PbPb signal shape is modeled using a triple Gaussian function, again with a common mean value but different widths. The widths and the relative areas of the Gaussian functions are obtained from the fit to simulated signal. A Chebyshev polynomial function is used to model the combinatorial background underlying the $\PcgLp$ signal including the background due to incorrect mass assignments for both $\Pp\Pp$ and \PbPb collisions. The order of the Chebyshev polynomial function for the nominal background shape is determined by a likelihood ratio test~\cite{LLR}.

As the event multiplicities for $\Pp\Pp$ and \PbPb collisions are substantially different, the selection criteria were optimized separately. The selection criteria optimization procedure mostly follows the one described in Ref.~\cite{HIN18009}, except that a gradient-boosted decision tree (BDTG) is chosen as the classification method. The variables used in the training are: the $\chi^2$ probability of vertex fit to the daughter tracks of \PcgLp, the angle between the \PcgLp momentum and the vector connecting the production and decay vertices, the distance between the two vertices, and the ratio of the daughter track \pt to that of the \PcgLp candidate. The BDTG optimization is used for the full \pt range 3--30\GeVc for $\Pp\Pp$ collisions and in the range 6--10\GeVc for \PbPb collisions. In the \pt range 10--40\GeVc for \PbPb collisions, the selection criteria are optimized for individual variables, since the BDTG training leads to irregular shapes in the side band region of the mass distribution. When performing the signal extraction, the widths and relative yields for the multiple Gaussian functions are fixed to the values found using MC simulations. To account for the difference in the widths of the Gaussian functions in the simulation and data, each width of the multiple Gaussian function is multiplied by an additional common scaling factor, obtained by fitting the mass distribution. The mean of the signal function is treated as a fit parameter. The pull distributions (defined here as the ratio of the difference between the data and the model fit value to the uncertainty in data) show the behavior of a random variable that is consistent with a Gaussian distribution with zero mean and unit width, which indicates the absence of bias in the fitting procedure.

\begin{figure}[!htb]
\centering
\includegraphics[scale=0.35]{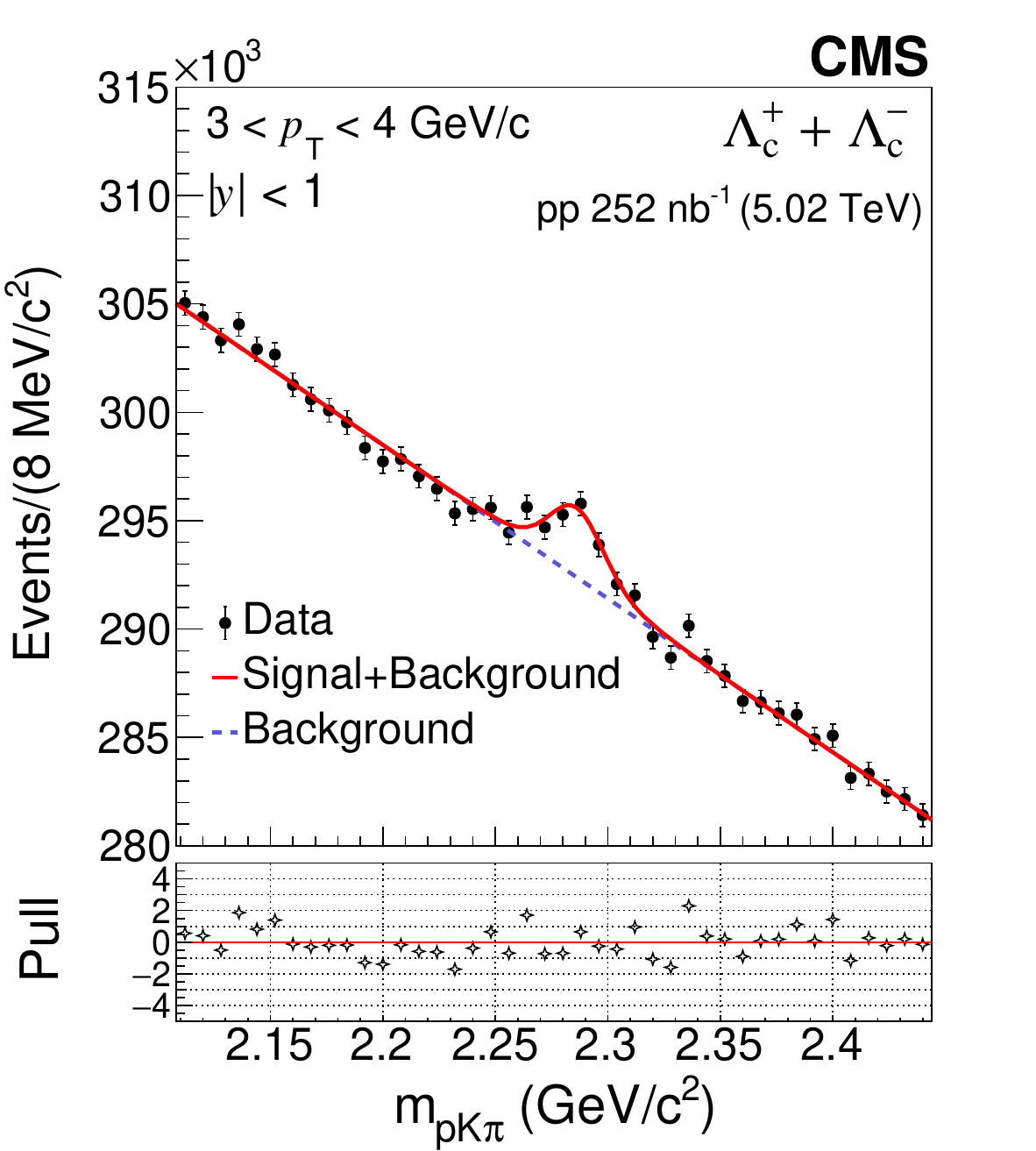}
\includegraphics[scale=0.35]{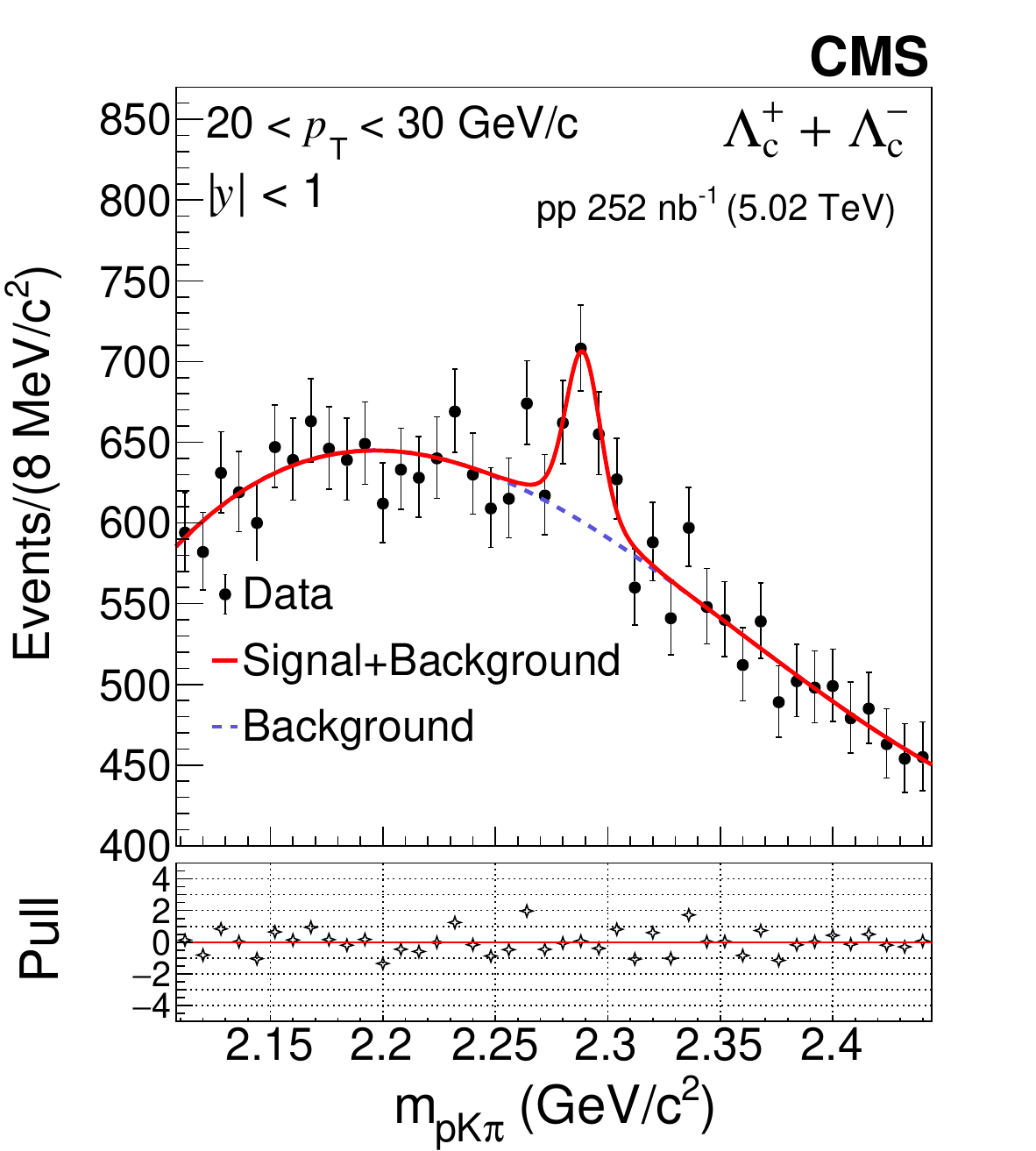}
\caption{The number of reconstructed \PcgLp candidates per 8\MeVcc invariant mass in $\Pp\Pp$ collisions for $\pt=3\text{--}4$ (left) and 20--30\GeVc (right). The solid line represents the fit to the data and the dashed line represents the fit to the background. The lower panels show the pulls, obtained as the difference between the data points and the fit result, divided by the uncertainty in data.}\label{pp_default}
\end{figure}

\begin{figure}[!htb]
\centering
\includegraphics[width=0.325\textwidth]{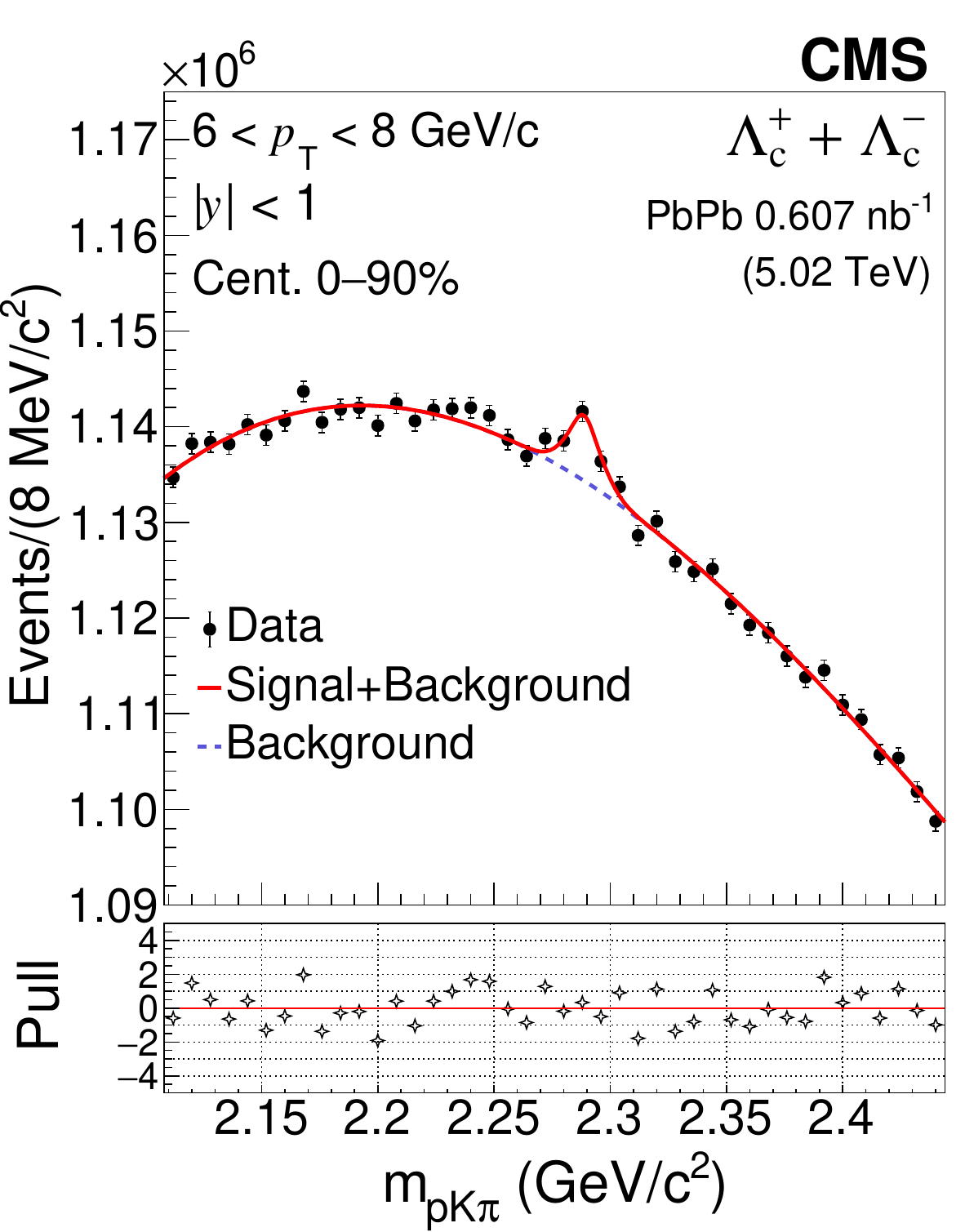}
\includegraphics[width=0.325\textwidth]{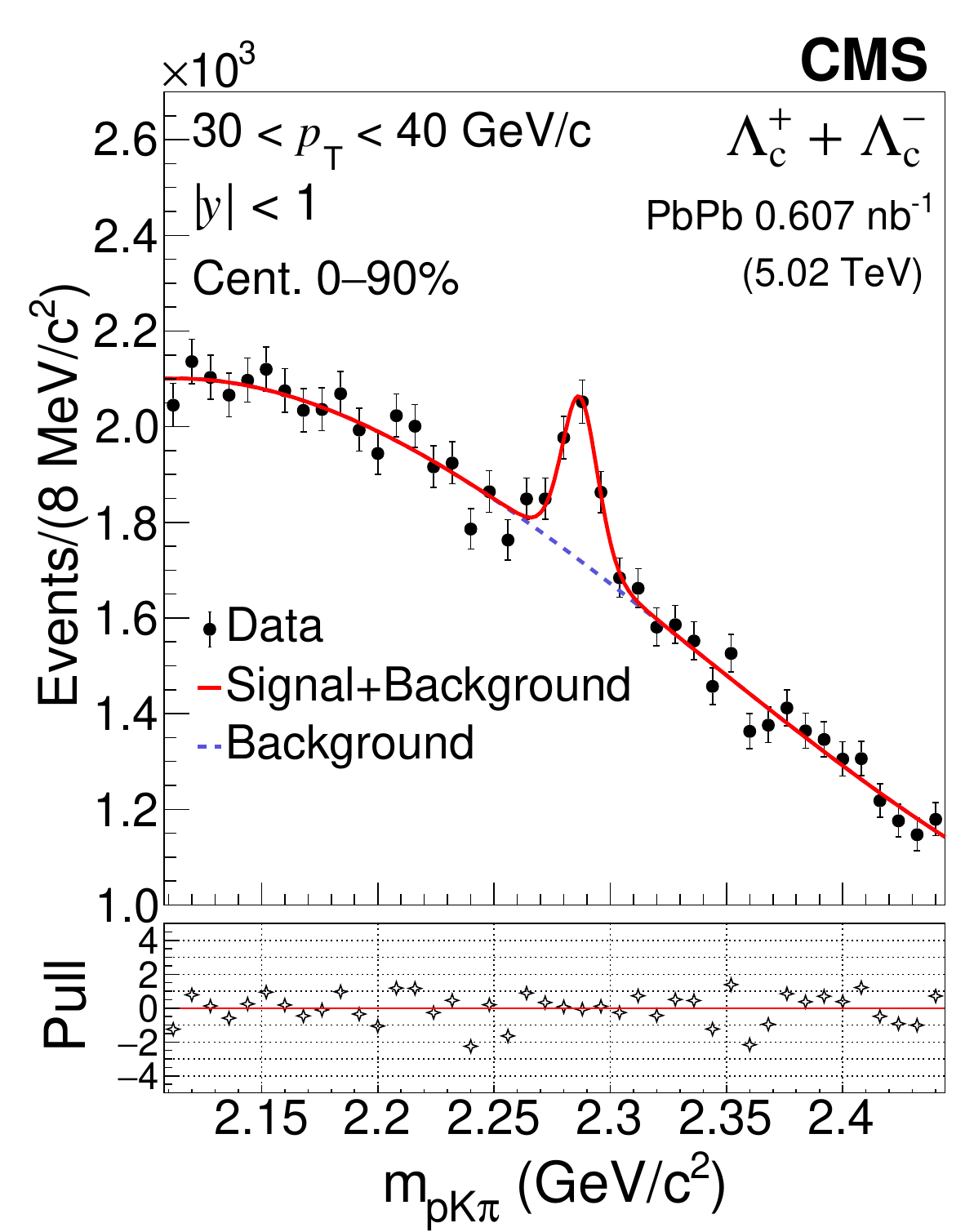}
\includegraphics[width=0.325\textwidth]{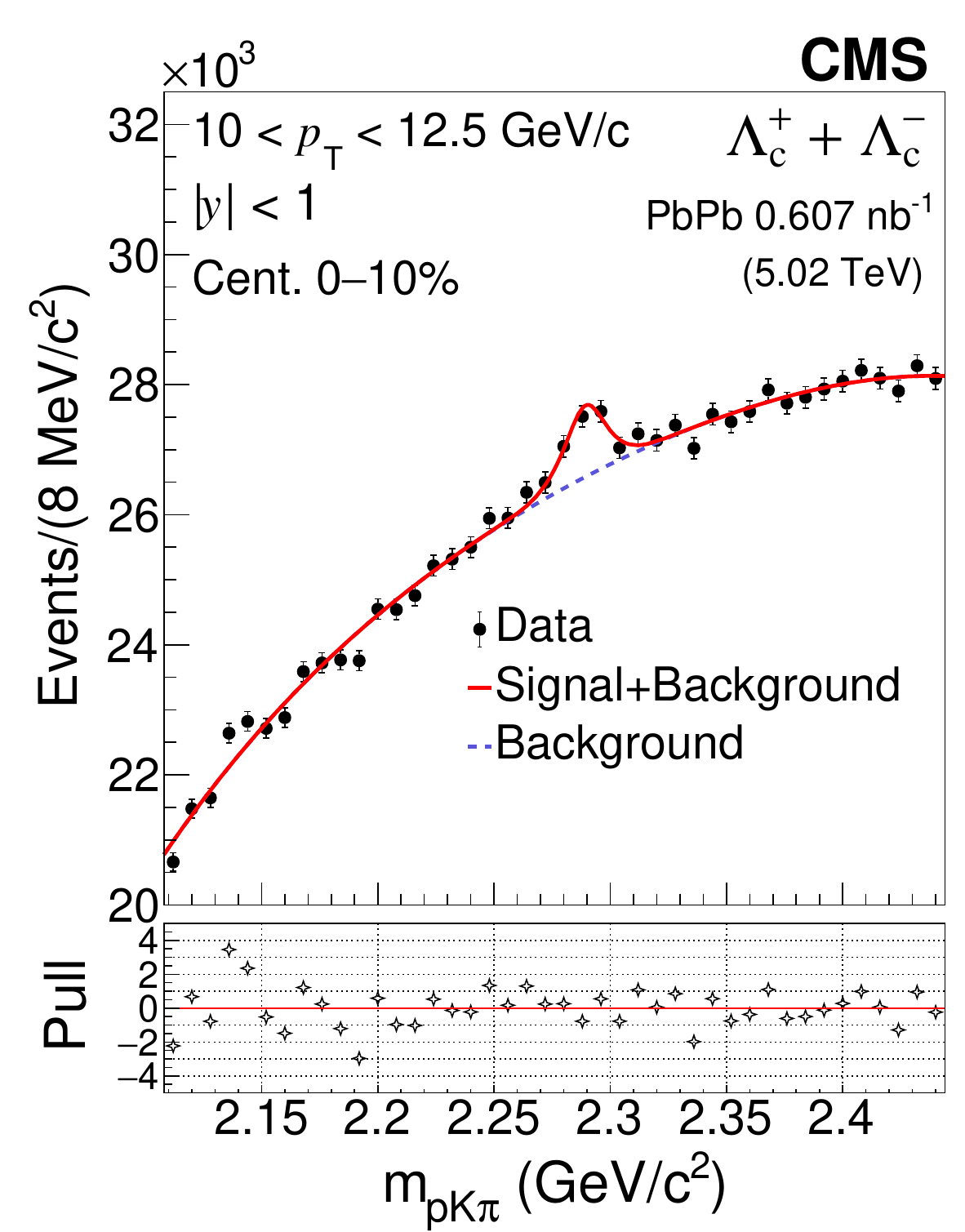}
\caption{The number of reconstructed \PcgLp candidates per $8\MeVcc$ invariant mass in \PbPb collisions for $\pt=6\text{--}8$ (left) and 30--40\GeVc (middle) in the 0--90\% centrality bin, and for $\pt=10\text{--}12.5\GeVc$ in the 0--10\% (right) centrality bin. The solid line represents the fit to the data and the dashed line represents the contribution from the combinatorial background. The lower panels show the pulls, obtained as the difference between the data points and the fit result, divided by the uncertainty in data.}\label{PbPb_default}
\end{figure}

Figure~\ref{pp_default} shows the invariant mass distribution of \PcgLp candidates in $\Pp\Pp$ collisions for $\pt=3\text{--}4$ (left) and 20--30\GeVc (right). Figure~\ref{PbPb_default} displays the invariant mass distribution of \PcgLp candidates for \PbPb collisions with $\pt=6\text{--}8$ (left) and 30--40\GeVc (middle) in the 0--90\% centrality bin, and for $\pt=10\text{--}12.5\GeVc$ (right) in the 0--10\% centrality bin. The solid line represents the fit to the data and the dashed line represents the contribution from the combinatorial background. For $\Pp\Pp$ collisions, a clear $\PcgLp$ signal peak is observed between $\pt=3\text{--}4$ and 20--30\GeVc. For \PbPb collisions, a significant signal is observed between $\pt=6\text{--}8$ and 30--40\GeVc for the 0--90\% centrality bin. For the 0--10\% centrality bin, the lowest \pt bin with a clearly resolved signal peak is 10--12.5\GeVc.

\section{Prompt signal fraction}
\label{sec:DCAfit}
The distribution of the distance of closest approach (DCA) of the inclusive $\PcgLp$ particles is used to obtain the prompt \PcgLp fraction $(f_{\text{prompt}})$. For \PbPb collisions, the DCA is defined as the closest three-dimensional distance between the $\PcgLp$ particle and the primary vertex. In case of a single event containing tracks from multiple collisions, the DCA can be distorted when the $\PcgLp$ does not originate from the primary vertex. To minimize this effect, the DCA for $\Pp\Pp$ collisions is taken as the closest two-dimensional distance in the transverse plane between the trajectory of the \PcgLp particle and the beamline, taking advantage of the small transverse beam size. Compared to that for prompt \PcgLp particles, the nonprompt \PcgLp DCA distribution, which results from \cPqb hadron decays, is wider. Templates of prompt and nonprompt \PcgLp distributions based on MC signal simulation are used to fit the data with the relative yield of the two components as the fit parameter. The fits are done with different DCA resolutions applied to the MC template to accommodate the imperfect description of the data by the MC simulations. The result of the fit with the minimum $\chi^2$ is chosen to be the nominal value of the prompt $\PcgLp$ fraction. The prompt fractions for $\Pp\Pp$ and \PbPb collisions remain relatively constant over the measured \pt and centrality (for \PbPb collisions) ranges, with values between 0.79--0.88 and 0.78--0.99, respectively. An exception occurs for PbPb collisions for the analysis bin corresponding to $\pt=10\text{--}12.5\GeVc$ and 0--10\% centrality. In this bin, the prompt ratio is 0.42 as a result of the tighter selection criteria. It is not feasible to obtain the DCA distribution from data for $\pt<4$ and ${>}15\GeVc$ in $\Pp\Pp$ collisions because of the small signal-to-background ratio and the limited number of events, respectively. For $\pt<4\GeVc$, the value of the neighboring \pt bin is used as the nominal value of the prompt \PcgLp fraction (0.85) and for individual bins with $\pt>15\GeVc$, the \pt bins are combined and the DCA fit is performed to obtain the nominal value (0.79).

Model predictions are obtained using $\PYTHIA 8$ with CR2~\cite{CRmodes} activated and fixed-order next-to-leading logarithmic (FONLL) calculations~\cite{FONLL}. The FONLL calculation does not include the nonprompt \PcgLp cross section. The nonprompt component from FONLL is obtained by weighting the nonprompt \PcgLp cross section from $\PYTHIA 8$ with the ratio of the \pt-differential \cPqb hadron cross sections obtained from FONLL to that from $\PYTHIA 8$. The $\PYTHIA 8$ prediction overestimates the observed \cPqb hadron cross section over the full \pt range~\cite{bhadron_data_comp}, while the FONLL prediction is much closer to the data. Therefore, the FONLL calculation is used to determine the systematic uncertainty related to the prompt fraction for $\pt<4\GeVc$, as discussed in Section~\ref{sec:sys}.

\section{Acceptance and efficiency}
The product of the acceptance and efficiency ($A\epsilon$) is obtained from the MC simulation as a fraction in which the numerator is the number of reconstructed prompt \PcgLp candidates that pass the signal selection criteria and the denominator is the number of generated prompt \PcgLp baryons in the considered \pt bin and with $\absy<1$. Since there are four sub-channels in the \Lcdecay decay, $A\epsilon$ is the weighted average of the four sub-channels according to their branching fractions. There is a slight discrepancy between the observed and simulated collision vertex distribution along the beam direction (\Vz) for both $\Pp\Pp$ and \PbPb collisions. A small discrepancy also exists with the centrality distribution of the events in \PbPb collisions. The discrepancy in the \Vz distribution is addressed by a weight factor that is the ratio of the \Vz distribution of data to that of the MC simulation. The discrepancy in the centrality distributions is addressed by a weight factor given by the ratio of the centrality distribution of data to that of the MC simulation multiplied with the average number of NN binary collisions ($\langle \Ncoll \rangle$). The values of $\langle \Ncoll \rangle$ are calculated using a MC Glauber model~\cite{Glauber}. The \pt spectrum of the generated \PcgLp baryons in the MC simulation is also weighted to match the data for both $\Pp\Pp$ and \PbPb collisions. This weighting procedure is done iteratively until sequential values of $A\epsilon$ converge. The distributions of variables used for the selection criteria optimization are shifted in the MC simulation so that the probability of having a \PcgLp candidate with a value equal to or lower than the nominal value in the data distribution is equivalent to the probability of having a candidate with a value equal to or lower than the adjusted value in the original MC distribution, before the $A\epsilon$ value is calculated. 

\begin{figure}[h]
    \centering
 \includegraphics[width=0.6\textwidth]{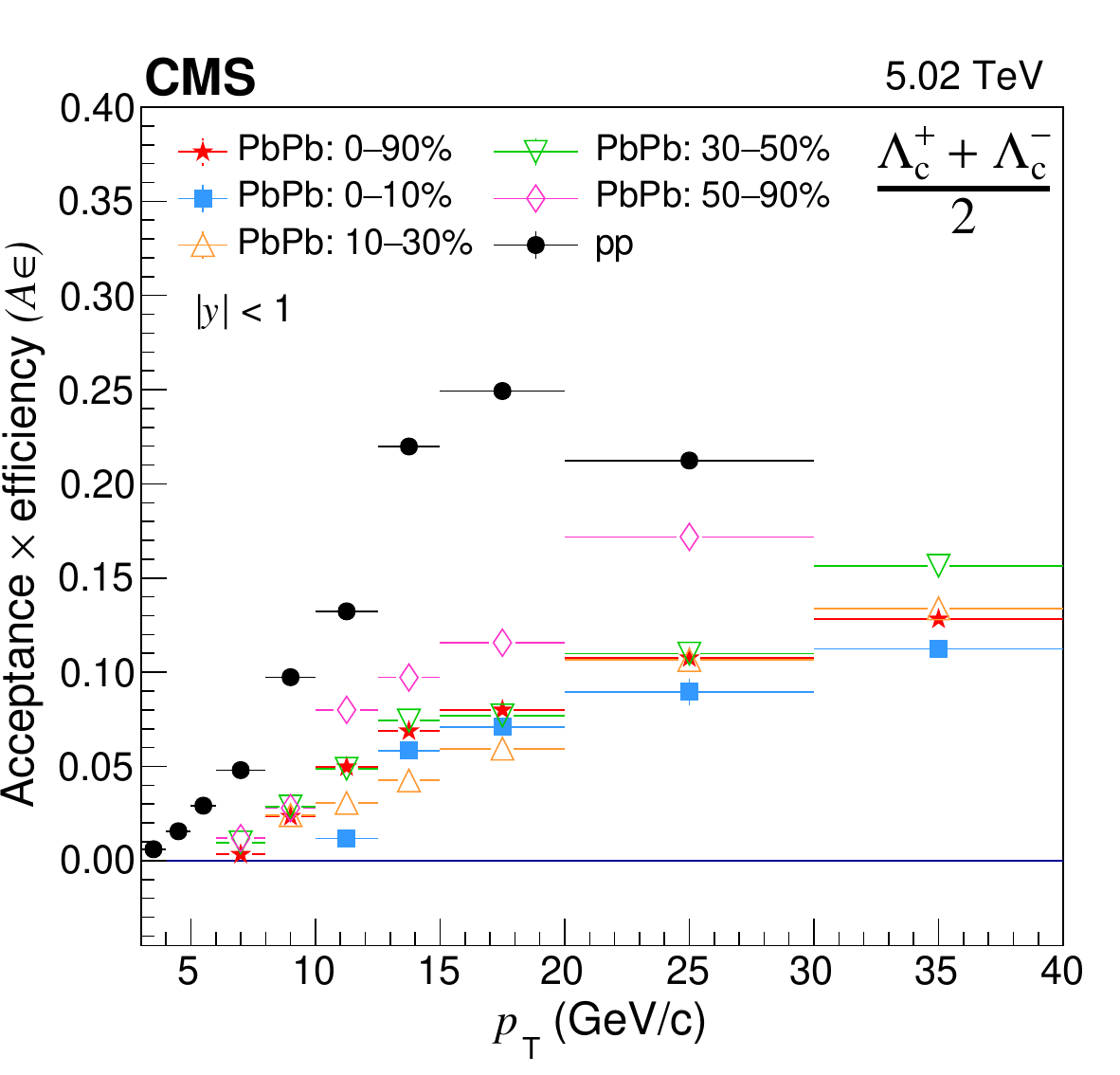}
 \caption{The product of acceptance and efficiency ($A\epsilon$) as a function of \pt for prompt $\PcgLp$ in $\Pp\Pp$ and \PbPb collisions. The closed circles represent the value for $\Pp\Pp$. The $A\epsilon$ values for \PbPb collisions in centrality bins 0--90, 0--10, 10--30, 30--50 and 50--90\% are represented by symbols of star, square, triangle, inverted triangle, and diamond, respectively. The horizontal error bars represent the bin widths.}
 \label{accp_pp_PbPb}
\end{figure}

Figure~\ref{accp_pp_PbPb} displays the $A\epsilon$ value as a function of \pt for prompt \PcgLp in $\Pp\Pp$ and \PbPb collisions. For the same \pt value in \PbPb collisions, the $A\epsilon$ value decreases toward more central collisions as a result of tighter selection criteria. Within the same centrality range, the $A\epsilon$ values increase from low to high \pt, mostly because of the longer average decay length of high \pt particles. The $A\epsilon$ values for $\Pp\Pp$ collisions are significantly higher than those for \PbPb collisions because of a smaller combinatorial background allowing for more relaxed selection criteria. The decrease in the $A\epsilon$ values for $\pt>20\GeVc$ in $\Pp\Pp$ collisions results from tighter selection criteria.

\section{Sources of systematic uncertainty}
\label{sec:sys}
The systematic uncertainties come from the extraction of the raw signal yield, the ability of the MC simulation to reproduce the data when varying selection criteria and in generating \pt spectra, the determination of prompt fraction from the raw signal yield, the branching fraction of the decay mode, and the integrated luminosity. Unless otherwise indicated, systematic uncertainties are combined by adding the individual contributions in quadrature. The systematic uncertainties mentioned here include the uncertainties for each \pt bin for $\Pp\Pp$ collisions and the uncertainties for each \pt and centrality bin for \PbPb collisions.

The systematic uncertainty in the raw signal yields is obtained by varying the modeling functions that are used for the signal and background contributions, as well as the fit range. The background function is changed from the default order to the next order Chebyshev polynomial, with the difference in yield between the alternative and the default fit functions taken as the systematic uncertainty. The uncertainties lie between 1--20\% for $\Pp\Pp$ collisions, and up to 34\% for \PbPb collisions. The signal yield uncertainty is largest for $\Pp\Pp$ collisions with $\pt=3\text{--}4\GeVc$ and for 0--10\% centrality \PbPb collisions with $\pt=12.5\text{--}15\GeVc$. The systematic uncertainty from the signal shape is estimated by changing from the default to an alternative signal fit function while keeping the background polynomial function the same as for the default fit. As alternative signal fit functions, three and two Gaussians are used for $\Pp\Pp$ and \PbPb collisions, respectively. The difference in the signal yield obtained with the default and alternative functions is taken as the systematic uncertainty, which is less than 6\% and 7\% for $\Pp\Pp$ and \PbPb collisions, respectively. The systematic uncertainty from the fit range is estimated by changing the fit range from the default mass range 2.11--2.45\GeVcc to alternative mass ranges of 2.17--2.37 and 2.15--2.41\GeVcc, for both $\Pp\Pp$ and \PbPb collisions. This leads to uncertainties within 2--26\% and 1--24\% for $\Pp\Pp$ and \PbPb collisions, respectively. The uncertainty is largest for $\Pp\Pp$ collisions with $\pt=4\text{--}5\GeVc$ and for 0--10\% centrality \PbPb collisions with $\pt=12.5\text{--}15\GeVc$.

The uncertainties associated with a particular selection criterion, such as the BDTG cut value, are determined from a double ratio defined as
\begin{linenomath}
\begin{equation}
    \mathcal{DR} =\left.\frac{N_\text{Data}\text{(varied)}}{N_\text{Data}\text{(nominal)}}\middle/\frac{N_\text{MC}\text{(varied)}}{N_\text{MC}\text{(nominal)}}\right.,
\end{equation}
\end{linenomath}
where $N_\text{Data}\text{(nominal)}$ and $N_\text{Data}\text{(varied)}$ are the yields obtained from data using the default and alternative selection criteria, respectively, and $N_\text{MC}\text{(nominal)}$ and $N_\text{MC}\text{(varied)}$ are the corresponding yields from the simulated events. The distribution of each selection criterion from MC is shifted so that the probability of a particular value for the \PcgLp candidate is the same in the MC simulation as in the data. The associated systematic uncertainty is then based on the sensitivity of the $\mathcal{DR}$ for a particular selection criterion to changes in this criterion~\cite{HIN18009}. For $\Pp\Pp$ and \PbPb collisions, the uncertainties are found to be less than 6\% and within 5--20\%, respectively. The uncertainty is largest for 0--90\% centrality \PbPb collisions with $\pt=30\text{--}40\GeVc$.

The difference between the experimental and MC \PcgLp \pt distributions leads to an uncertainty when estimating $A\epsilon$. For both $\Pp\Pp$ and \PbPb collisions, this resulting uncertainty is taken as the difference in the estimated $A\epsilon$ value when using the experimental \pt distribution and the corresponding $\PYTHIA 8$ distribution with CR2 turned on. The resulting systematic uncertainty is up to 4\% and 2\% for $\Pp\Pp$ and \PbPb collisions, respectively.

The accuracy of the calculated $A\epsilon$ value also depends on knowledge of the resonant substructure of the $\Pp\PKm\Pgpp$ decay mode~\cite{Workman:2022ynf}. The calculation of $A\epsilon$ uses the appropriately weighted sum of the four known sub-channels. Pseudo-experiments are used to determine the distribution of $A\epsilon$ by randomly sampling the value of the branching fraction based on the corresponding uncertainty for each sub-channel. The systematic uncertainty is obtained as the standard deviation of a Gaussian fit to the $A\epsilon$ distribution and varies from 8--9\% for $\Pp\Pp$ and \PbPb events.

The uncertainty of the single-track reconstruction efficiency is 2\% for $\Pp\Pp$ collisions and 5\% for \PbPb collisions~\cite{HIN16011}. Since there are three tracks in the $\PcgLp$ decay, the corresponding uncertainties on the measured \pt spectra are 6 and 15\% for $\Pp\Pp$ and \PbPb collisions, respectively, assuming that the track efficiencies are correlated. The prompt \PDz cross section measured using $\Pp\Pp$ collision data collected in 2015~\cite{HIN16001} is used to calculate the \LcDratio production ratio in this analysis. Since the uncertainty in the single track reconstruction for that analysis is 4\%, we use this larger value when calculating the \LcDratio uncertainty for $\Pp\Pp$ collisions. The uncertainties of reconstructing kaon and pion daughter tracks from \PDz and \PcgLp cancel, leading to the uncertainties from this source to be 4\% in $\Pp\Pp$ collisions and 5\% in \PbPb collisions as a result of the proton daughter track. 

The uncertainties arising from the prompt fraction determination for most \pt bins are estimated by comparing the value obtained from the DCA distribution and the DCA significance distribution of the signal. The DCA significance is defined as the ratio of DCA to DCA uncertainty. The consistency between the data and MC in DCA significance is better than in DCA because the discrepancy between MC and data for the DCA and the DCA uncertainty partially cancel. The difference in the prompt fraction obtained from the two methods is treated as the systematic uncertainty of the prompt fraction. For $\pt<4\GeVc$, the systematic uncertainty is calculated as the difference between the nominal value and FONLL predictions. For $\pt>15\GeVc$, the difference between the prompt fraction obtained from DCA and the DCA significance for $12.5<\pt<30\GeVc$ is taken as the systematic uncertainty. The systematic uncertainty from this source is within 6--16\% and up to 43\% for $\Pp\Pp$ and \PbPb collisions, respectively. The uncertainty is largest for 30--50\% centrality \PbPb collisions with $\pt=30\text{--}40\GeVc$.

The overall \Lcdecay branching fraction uncertainty is 5\%~\cite{Workman:2022ynf}. The uncertainties in the integrated luminosity for $\Pp\Pp$ collisions and the MB selection efficiency for \PbPb collisions are 2 and 1\%, respectively. The uncertainties in the mean nuclear overlap function, $\langle \taa \rangle$, for \PbPb collisions are between 2--5\%. The $\langle \taa \rangle$ value is used in the calculation of \RAA and the \taa-scaled \PcgLp yields for \PbPb collisions, as discussed in Section 8. The uncertainty based on the centrality calibration is obtained by varying the centrality bin edges by ${\pm}0.5\%$, and taking the maximum relative difference with the default signal yield. The systematic uncertainty associated with the centrality calibration in \PbPb collisions is up to 12\%. The uncertainty is largest for 30--50\% centrality \PbPb collisions with $\pt=15\text{--}20\GeVc$.

For the measurement of the \pt spectra, the uncertainties associated with the \Lcdecay branching fraction, luminosity, tracking efficiency, and MB selection are taken as global ones. Adding these systematic uncertainty contributions in quadrature yields the global uncertainties of 8.6 and 15.9\% for $\Pp\Pp$ and \PbPb collisions, respectively. When determining \RAA values, the uncertainties associated with the branching fraction and subresonant contributions cancel. The uncertainties associated with the \PDz yields from the yield extraction, selection criteria efficiency, and \pt shape, when calculating the \LcDratio production ratio, are obtained from Ref.~\cite{HIN16001}. The uncertainties in tracking efficiency and in the branching fractions of \PcgLp and \PDz are included in the global systematic uncertainty for \LcDratio production ratio. The global uncertainty is 7\% for both $\Pp\Pp$ and \PbPb collisions. The sources of the systematic uncertainties are summarized in Table~\ref{sys_uncern_table}.

\begin{table}[]
\centering
\topcaption{Systematic uncertainties from different sources.}
\begin{tabular}{lcc}
\hline
Sources of systematic uncertainties             & \multicolumn{1}{c}{$\Pp\Pp$}      & \PbPb         \\  [\cmsTabSkip]
Signal extraction (background)           		& \multicolumn{1}{c}{1--20\%}  	    & ${<}34$\%   \\ 
Signal extraction (signal)               		& \multicolumn{1}{c}{${<}6$\%}      & ${<}7$\%      \\
Signal extraction (fit range)            		& \multicolumn{1}{c}{2--26\%}       & 1--24\%       \\
Selection efficiency                 			& \multicolumn{1}{c}{${<}6$\%}      & 5--20\%       \\ 
\taa            								& \multicolumn{1}{c}{\NA  }			& 2--5\%		\\
MC \pt shape                       			    & \multicolumn{1}{c}{${<}4$\%}      & ${<}2$\%      \\
$A \epsilon_\text{reco}$ 						& \multicolumn{2}{c}{8--9\%}                        \\ 
Tracking efficiency                  			& \multicolumn{1}{c}{6\%} 		    & 15\%   		\\ 
Prompt fraction                      			& \multicolumn{1}{c}{6--16\%}       & ${<}43$\%     \\ 
Branching fraction                   			& \multicolumn{2}{c}{5\%}        	    			\\ 
Integrated luminosity							& \multicolumn{1}{c}{2\%}			& \NA			\\
Minimum bias selection efficiency       			& \multicolumn{1}{c}{\NA }     		& 1\%           \\ 
Centrality calibration               			& \multicolumn{1}{c}{\NA }          & ${<}12$\%     \\ \hline
\end{tabular}
\label{sys_uncern_table}
\end{table}

\section{Results}
\label{sec:results}
\begin{figure}[htb]\centering
 \includegraphics[width=0.6\textwidth]{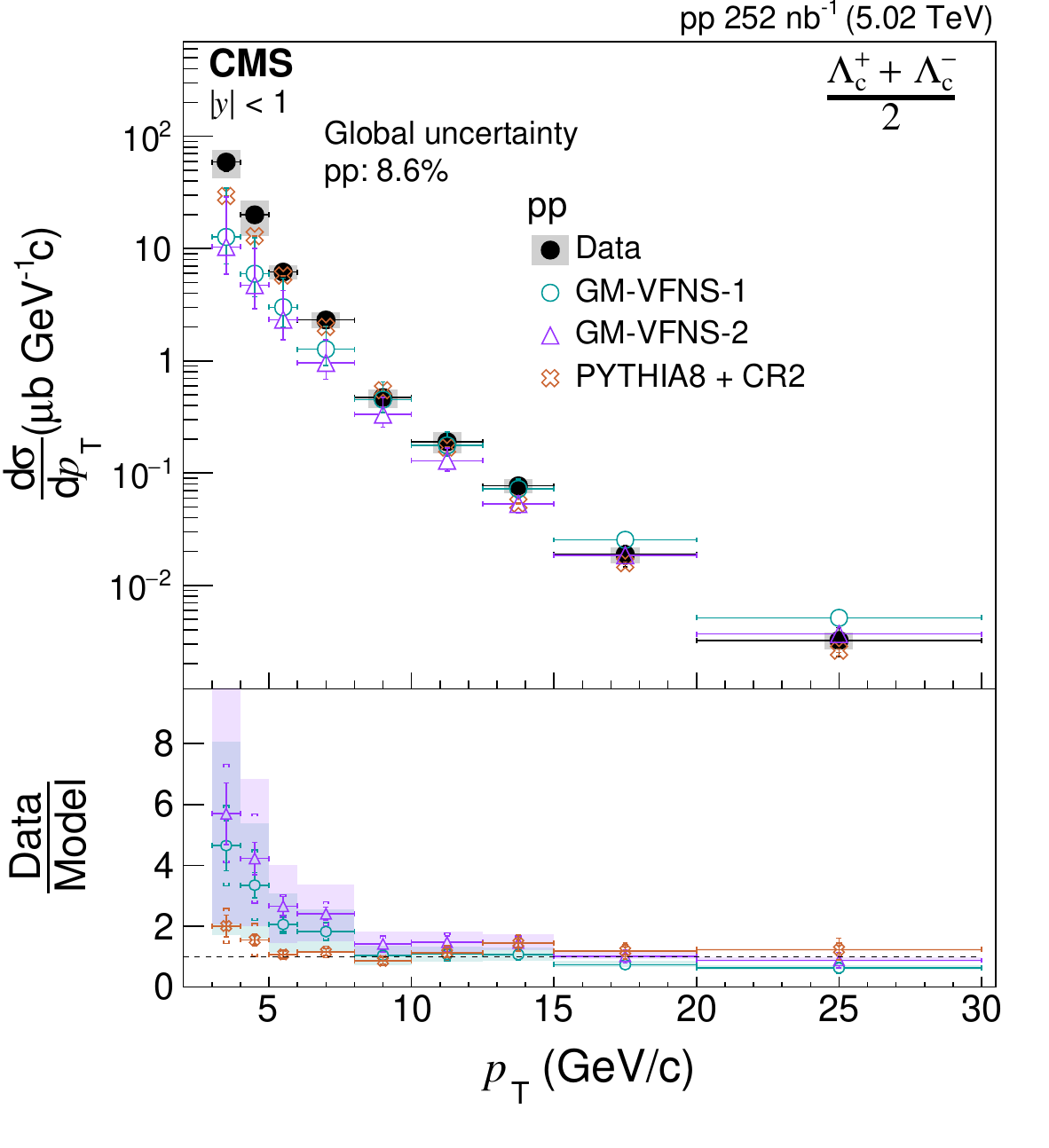}
 \caption{The \pt-differential cross sections for prompt $\PcgLp$ production in $\Pp\Pp$ collisions. 
Predictions for $\Pp\Pp$ collisions are displayed for $\PYTHIA 8$ with CR2 (open crosses), GM-VFNS implementing fragmentation functions that are fitted to the OPAL data, only (open circles labeled GM-VFNS-1), and fitted to both OPAL and Belle data (open triangles labeled GM-VFNS-2). The GM-VFNS model calculations are for inclusive \PcgLp production. The horizontal error bars represent the bin widths. The vertical lines in the data points represent the statistical uncertainties and the shaded boxes represent the systematic uncertainties. The vertical lines in the model points represent the GM-VFNS uncertainties. The lower panel shows the data-to-prediction ratio for $\Pp\Pp$ collisions with error bars and brackets corresponding to the statistical and total uncertainties in the data, respectively. The global fit uncertainty of 8.6\% is not shown in the plot. The shaded boxes in the lower panel represent the GM-VFNS uncertainties.}
\label{pp_spectrum}
\end{figure}

The \pt-differential cross section for $\Pp\Pp$ collisions is defined as:
\begin{linenomath}
\begin{equation}
\label{eqn:pp_cross}
   \left.\frac{\rd\sigma^{\PcgLp}_{\Pp\Pp}}{\rd\pt}\right|_{\abs{y}<1} = \frac{f_{\text{prompt}}}{2 \lumi \Delta\pt \mathcal{B}}\left.\frac{N_{\Pp\Pp}^{\PcgLp}|_{\abs{y}<1}}{A\epsilon}\right.,
\end{equation}
\end{linenomath}
where $N_{\Pp\Pp}^{\PcgLp}|_{\absy<1}$ is the \PcgLp raw yield extracted in a \pt interval with $\absy<1$, \lumi is the luminosity of ZB events, $\Delta \pt$ is the width of the \pt interval, $f_{\text{prompt}}$ is the prompt \PcgLp fraction, $\mathcal{B}$ is the branching fraction of the decay channel, and the factor of $1/2$ accounts for measurements of yields for both particles and antiparticles, but quoting the cross section for only particles. Figure~\ref{pp_spectrum} shows the \pt-differential cross sections for prompt \PcgLp baryon production in $\Pp\Pp$ collisions with $\absy<1$. The global 8.6\% normalization uncertainty for the $\Pp\Pp$ results is not included in the boxes representing the systematic uncertainties for each data point. The data are well described by $\PYTHIA 8$ after activating CR2 in which the final partons in the string fragmentation are considered to be color connected in such a way that the total string length becomes as short as possible~\cite{CRmodes}. The cross sections are also compared to the GM-VFNS perturbative QCD calculations~\cite{gmvfns1, gmvfns2}, which utilize fragmentation functions obtained by fitting data from the OPAL and Belle Collaborations~\cite{OPAL:LC_Study, Belle:LC_Study}. The GM-VFNS models, implementing fragmentation functions that are fitted to the OPAL data, only, and to both the OPAL and the Belle data are shown as GM-VFNS-1 and GM-VFNS-2, respectively. The error bars on the GM-VFNS predictions account for the renormalization scale uncertainty. The GM-VFNS predictions, which include both prompt and nonprompt baryon production, are systematically below our data for $\pt<8\GeVc$, similar to the difference found by previous ALICE~\cite{ALICE_pp} and CMS measurements~\cite{HIN18009}. The DCA distribution fit of the data shows that ${\sim}20$\% of the generated \PcgLp baryons arise from \cPqb hadrons in the \pt range 3--30\GeVc. Therefore, accounting for the effects of nonprompt \PcgLp production slightly enhances the disagreement with the data. The discrepancies in the cross sections and the \LcDratio production ratio~\cite{HIN18009} between the data and the GM-VFNS calculations in $\Pp\Pp$ collisions indicate a breakdown of the universality of charm quark to hadron fragmentation functions.

\begin{figure}[htb]\centering
 \includegraphics[width=0.6\textwidth]{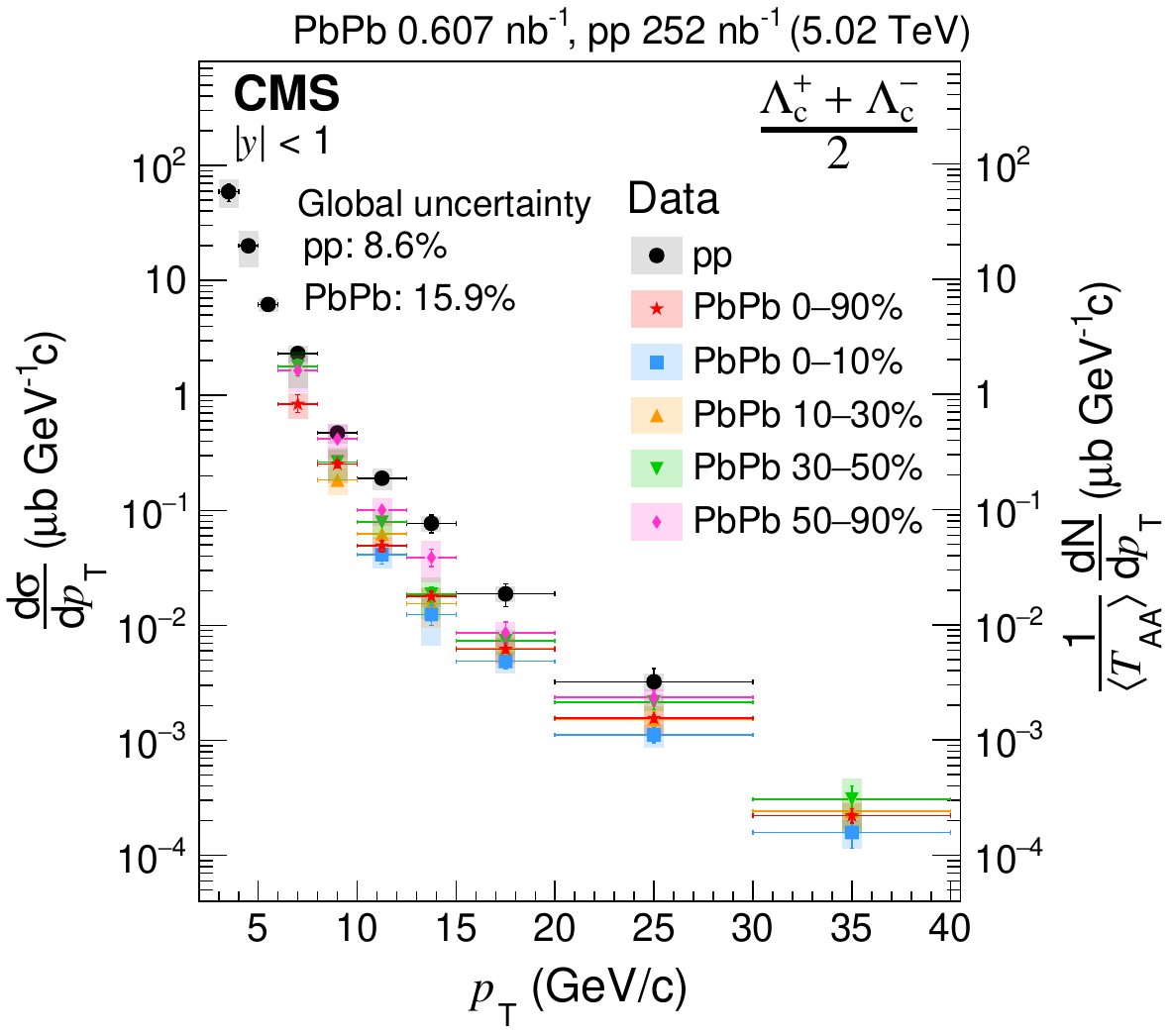}
 \caption{The \pt-differential cross sections for prompt $\PcgLp$ production in $\Pp\Pp$ collisions (circles) and the \taa-scaled yields for \PbPb collisions within centrality regions of 0--90 (stars), 0--10 (squares), 10--30 (triangles), 30--50 (inverted triangles) and 50--90\% (diamonds) in \PbPb collisions. The boxes and error bars represent the systematic and statistical uncertainties, respectively. The horizontal error bars represent the bin widths.
}\label{yield_PbPb}\end{figure}

The mean nuclear overlap function, $\langle \taa \rangle$, is equal to the average number of NN binary collisions ($\langle \Ncoll \rangle$) divided by the $\Pp\Pp$ inelastic cross section and can be interpreted as the NN-equivalent integrated luminosity per heavy ion collision. The $\taa$-scaled yields are defined as:
\begin{linenomath}
\begin{equation}
       \frac{1}{\langle \taa \rangle}\left.\frac{\rd N^{\PcgLp}_{\PbPb}}{\rd\pt}\right|_{\abs{y}<1} = \frac{f_{\text{prompt}}}{\langle \taa \rangle}\frac{1}{2N_{\text{events}}\Delta\pt\mathcal{B}}\left.\frac{N_{\PbPb}^{\PcgLp}|_{\abs{y}<1}}{A\epsilon}\right.,
\end{equation}
\end{linenomath}
where $N_{\text{events}}$ is the number of MB events. The other variables refer to the same physical quantities as in Eq.~(\ref{eqn:pp_cross}). Figure~\ref{yield_PbPb} displays the \taa-scaled yields for \PbPb collisions in centrality bins of 0--90, 0--10, 10--30, 30--50 and 50--90\% with $\pt<40\GeVc$ and $\absy<1$. The lowest \pt values for the measurements vary with centrality as the number of events and the signal-to-background ratio change, with thresholds of 6\GeVc in the 0--90, 30--50 and 50--90\% centrality bins, 8\GeVc in the 10--30\% bin, and 10\GeVc in the 0--10\% bin. The 15.9\% normalization uncertainty for the \PbPb results is not included in the boxes representing the systematic uncertainties. The $\Pp\Pp$ measurement is included in the figure for comparison. Compared to the \PcgLp \pt-differential cross sections for $\Pp\Pp$ collisions, the $\taa$-scaled \PcgLp yields for \PbPb collisions are systematically lower, with the reduction becoming more significant toward central collisions. 

\begin{figure}[!htb]\centering
 \includegraphics[width=0.6\textwidth]{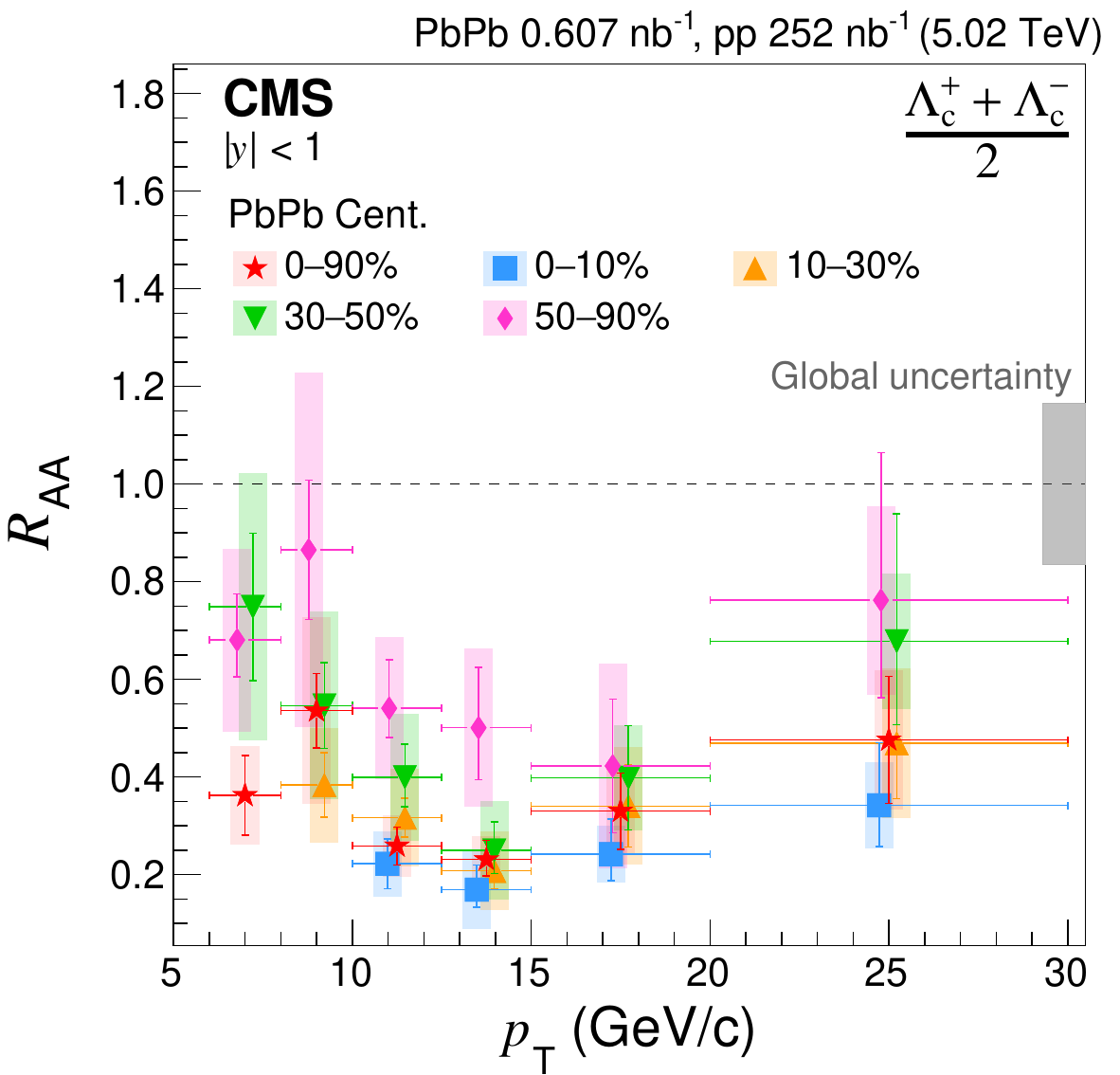}
 \caption{The nuclear modification factor $\RAA$ versus \pt for prompt $\PcgLp$ production in centrality regions of 0--90 (stars), 0--10 (squares), 10--30 (triangles), 30--50 (inverted triangles) and 50--90\% (diamonds) in \PbPb collisions. The mean position of the data points are shifted along the horizontal axis for clarity. The boxes and error bars represent the systematic and statistical uncertainties, respectively. The horizontal error bars represent the bin widths. The band at unity labeled global uncertainty includes the uncertainties for the luminosity of $\Pp\Pp$ collisions, the number of MB events in \PbPb collisions, and the tracking efficiency. The global uncertainty for $\RAA$ is 16.5\%.}
 \label{Raa_plot}\end{figure}

The \RAA value is used to quantify the suppression in production yield per NN collisions in \PbPb relative to $\Pp\Pp$ collisions. The \PcgLp \RAA value for \PbPb collisions is defined as
\begin{linenomath}
\begin{equation}
    \RAA(\pt) = \frac{1}{\langle \taa \rangle} \left.\frac{\rd N_\text{PbPb}^{\PcgLp}}{\rd\pt}\middle/\frac{\rd\sigma_{\Pp\Pp}^{\PcgLp}}{\rd\pt}\right..
\label{Raa_equation}
\end{equation}
\end{linenomath}
Figure~\ref{Raa_plot} shows the \RAA values for \PcgLp as a function of \pt for \PbPb collisions within centrality regions of 0--90, 0--10, 10--30, 30--50 and 50--90\% in $\absy<1$. A larger suppression factor in \PcgLp production per NN collision is observed for more central \PbPb collisions. Comparisons with the ALICE measurements~\cite{ALICE_PbPb2}, which extend to lower \pt values, for the 0--10 and 30--50\% centrality ranges are displayed in Fig.~\ref{app:Raa_plot_ALICE_comparison} (in App.~\ref{app:Figures}). The results suggest that the \PcgLp \RAA values are smallest near $\pt\approx14\GeVc$, with larger values found at both lower and higher \pt. The trend is, in general, consistent with the observation from other heavy flavor hadron measurements, although the \pt of the minimum $\RAA$ is different. For instance, the $\RAA$ reaches a minimum at $\pt\approx8\GeVc$ for nonprompt \PDz from \cPqb hadron decay and at $\pt\approx9\GeVc$ for prompt \PDz~\cite{HIN16016}.

\begin{figure}[!htb]
\centering
\includegraphics[width=0.48\textwidth]{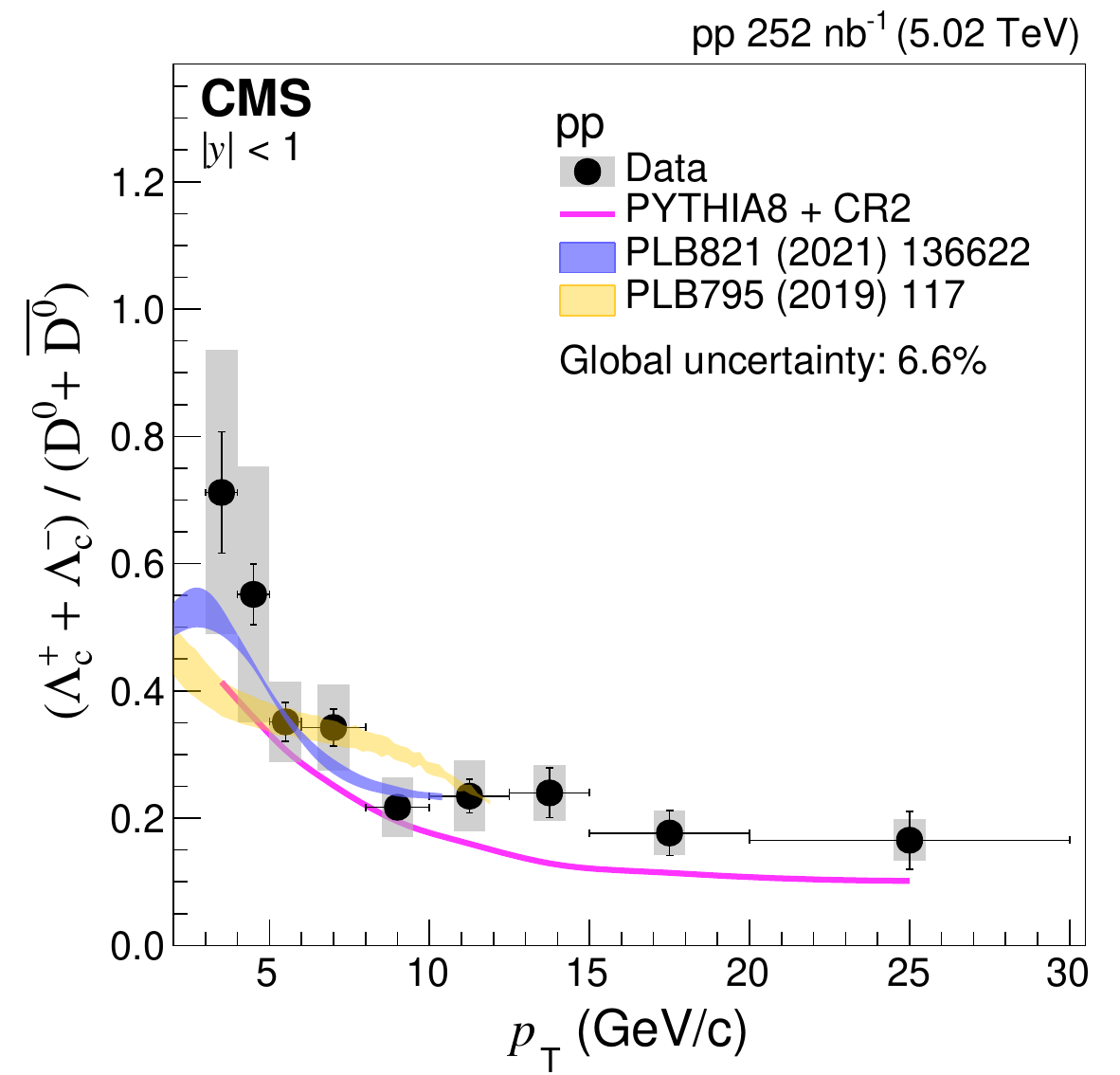}
\includegraphics[width=0.48\textwidth]{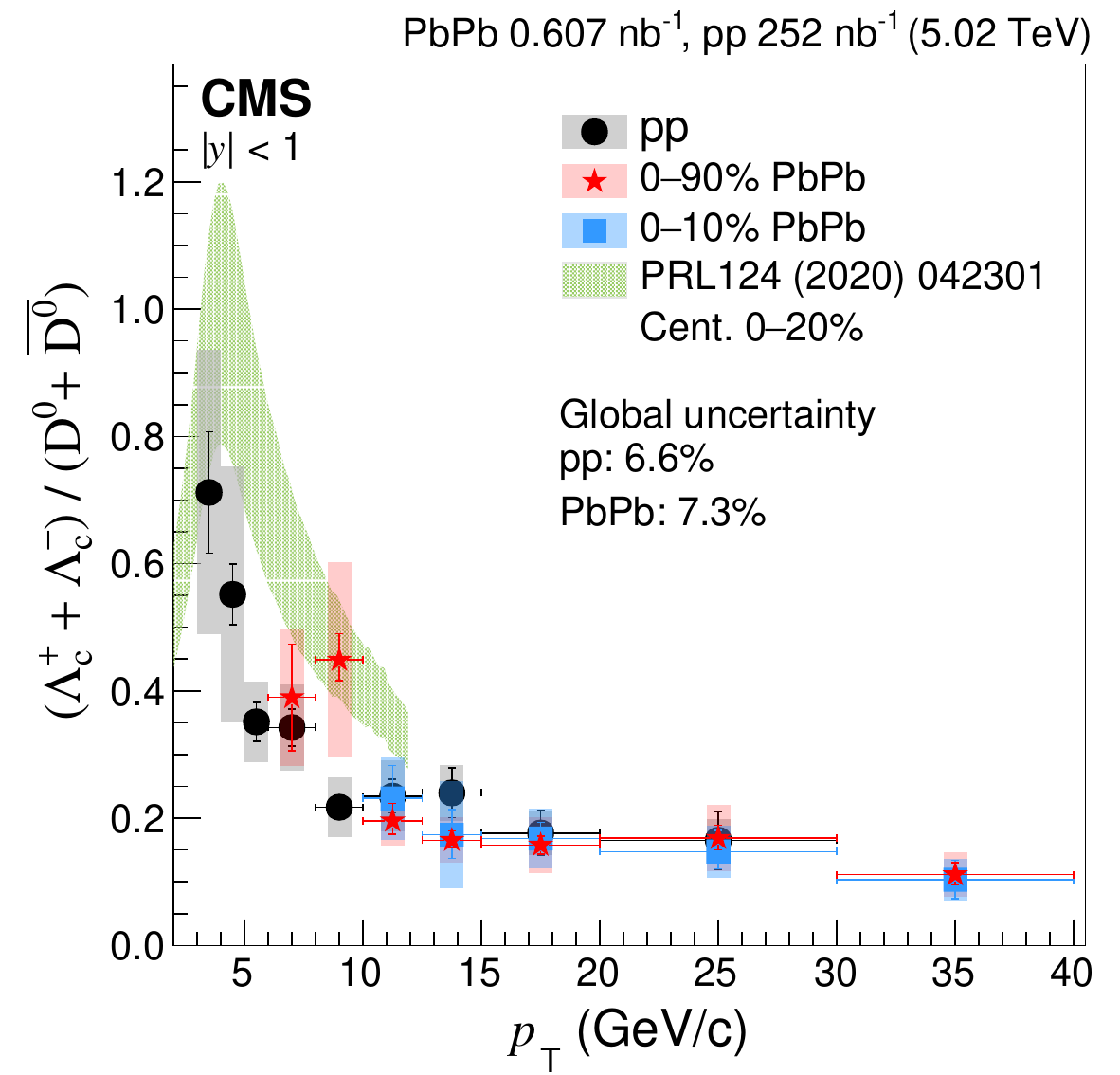}
\caption{The ratio of the production cross sections of prompt $\PcgLp$ to prompt $\PDz$ versus \pt from $\Pp\Pp$ collisions is represented by closed circles (left). The ratio for  0--90 (closed stars) and 0--10\% (closed squares) centrality classes of \PbPb collisions are compared to the $\Pp\Pp$ result (right). The boxes and error bars represent the systematic and statistical uncertainties, respectively. The horizontal error bars represent the bin widths. The 6.6 and 7.3\% normalization uncertainties in $\Pp\Pp$ and \PbPb collisions, respectively, are not included in the boxes representing the systematic uncertainties for each data point. Model calculations are displayed (see text for details).}
\label{LcD0_plot}
\end{figure}

The \LcDratio production ratio as a function of \pt for $\Pp\Pp$ collisions and in the 0--90 and 0--10\% centrality ranges for \PbPb collisions in $\absy<1$ are shown in Fig.~\ref{LcD0_plot}. The ratio is calculated using the prompt $\PcgLp$ measurements in this analysis and the previous CMS measurement for prompt \PDz mesons~\cite{HIN16001}. The purple curve in Fig.~\ref{LcD0_plot}(left) represents the prediction of $\PYTHIA 8$ with CR2~\cite{CRmodes}. This prediction is consistent with our measurement in $\Pp\Pp$ collisions for $\pt<10\GeVc$, and is systematically lower than observed for $10<\pt<30\GeVc$. Two model calculations of the prompt \LcDratio ratio are also compared to the $\Pp\Pp$ data. The dark blue band in Fig.~\ref{LcD0_plot}(left) represents the updated prediction from a model, labeled as PLB$821(2021)136622$ in the figure, that includes both coalescence and fragmentation processes in $\Pp\Pp$ collisions~\cite{Greco_pub_2}. The update consisted in improvements on a previous calculation~\cite{Greco_publication} that predicted a stronger dependence on \pt and underestimated the earlier CMS measurements~\cite{HIN18009}. The improvements include raising the underlying charm quark distribution to be closer to the upper limit of the FONLL prediction, and updating the charm fragmentation function to reproduce data for $\pt>10\GeVc$. The new calculation describes the data very well for $\pt<10\GeVc$. Another model~\cite{Rapp_ppresult}, labeled as PLB$795(2019)117$ in Fig.~\ref{LcD0_plot}(left), uses a statistical hadronization approach and explains the large \LcDratio production ratio as arising from \PcgLp baryons that are produced from the decay of excited charm baryon states not included in the particle data group (PDG) Ref.~\cite{Workman:2022ynf} and therefore not included in the hadronization simulation in $\PYTHIA 8$. This model calculation, shown by the light yellow band, provides a reasonable description of the data for $\pt<12\GeVc$.

The measurements of the \LcDratio ratio in 0--90 and 0--10\% \PbPb collisions with $\pt>10\GeVc$ are consistent with the $\Pp\Pp$ result. The \LcDratio ratios in $\Pp\Pp$ and \PbPb collisions approach the value found for $\Pep\Pem$ collisions~\cite{Eventune,HIN18009} in this higher-\pt region. This lack of enhancement suggests that there is no significant contribution from the coalescence process in this \pt region for \PbPb collisions. For $\pt<10\GeVc$ and 0--90\% centrality, the significance of the current measurement is not high enough to draw a definite conclusion. For $\pt>10\GeVc$ and with centrality in the 0--10\% range, the results are consistent with those of the ALICE Collaboration~\cite{ALICE_PbPb2}, as shown in Fig.~\ref{app:LcD0_plot_ALICE_comparison} (in App.~\ref{app:Figures}). The measurements of the \LcDratio production ratio in 0--10\% \PbPb collisions are also compared to a model calculation~\cite{Rapp_PbPb}, represented by the light green band in Fig.~\ref{LcD0_plot}(right). The model uses a four-momentum conserving recombination mechanism for baryons and applies space-momentum correlations between charm quarks and the hydrodynamical medium on an event-by-event basis. It also includes the decays of excited charm baryon states not included in Ref.~\cite{Workman:2022ynf}. Although the calculation is done for the 0--20\% centrality bin, the prediction for 0--10\% is almost identical and consistent with the data within uncertainties for $10<\pt<12.5\GeVc$. 

\section{Summary}
The differential cross section of prompt \PcgLp baryons as a function of transverse momentum (\pt) is presented for both proton-proton ($\Pp\Pp$) and lead-lead (\PbPb) collisions at a center-of-mass energy per nucleon pair of 5.02 TeV in the central rapidity region $\absy<1$. The measured \pt ranges are 3--30\GeVc and 6--40\GeVc  for the $\Pp\Pp$ and \PbPb collisions, respectively. The \pt range of the $\Pp\Pp$ and \PbPb results, together with the differential centrality study of the prompt \PcgLp production, significantly extend previous CMS results based on 2015 data that was obtained with a lower integrated luminosity. The \PcgLp baryon yields for $\Pp\Pp$ collisions are much higher than predicted by calculations with the general-mass variable-flavor-number scheme that use fragmentation functions obtained by fitting results from the OPAL and Belle Collaborations, indicating a breakdown of the universality of charm quark fragmentation functions. The nuclear modification factors, which correspond to the $\PcgLp$ yields divided by the $\Pp\Pp$ yields scaled up by the number of nucleon-nucleon collisions, have been measured in various centrality classes for the \PbPb collisions. The prompt \PcgLp production for \PbPb collisions is significantly suppressed compared to the $\Pp\Pp$ collision results. The suppression magnitude is larger in more central collisions and shows a hint of varying with the \PcgLp baryon \pt. This is consistent with the partonic energy loss affecting charm quarks in the quark-gluon plasma. A similar effect was previously observed for (charmed) \PDz mesons and many other hadrons. The \LcDratio production ratio is measured for $\Pp\Pp$ collisions with \pt in the range 3--30\GeVc, 0--90\% centrality \PbPb collisions with \pt in the range 6--40\GeVc, and 0--10\% centrality \PbPb collisions with \pt in the range 10--40\GeVc. Calculations based on the event generator $\PYTHIA 8$ for the \LcDratio production ratio with the inclusion of color reconnection (mode 2) in the hadronization step can describe the $\Pp\Pp$ data well for $\pt<10\GeVc$, but are systematically lower for $10<\pt<30\GeVc$. A model taking into account the contributions from the decays of excited charm baryons and a model involving both coalescence and fragmentation can also describe the \LcDratio production ratios in $\Pp\Pp$ collisions. For $\pt>10\GeVc$, the \LcDratio ratios for $\Pp\Pp$ and \PbPb collisions are consistent with each other and approach the value found for $\Pep\Pem$ collisions, suggesting that the coalescence process does not play a significant role in \PcgLp baryon production in this higher-\pt region.

\begin{acknowledgments}
We thank Denes Molnar for his aid in improving the efficiency of the analysis computations.

We congratulate our colleagues in the CERN accelerator departments for the excellent performance of the LHC and thank the technical and administrative staffs at CERN and at other CMS institutes for their contributions to the success of the CMS effort. In addition, we gratefully acknowledge the computing centers and personnel of the Worldwide LHC Computing Grid and other centers for delivering so effectively the computing infrastructure essential to our analyses. Finally, we acknowledge the enduring support for the construction and operation of the LHC, the CMS detector, and the supporting computing infrastructure provided by the following funding agencies: SC (Armenia), BMBWF and FWF (Austria); FNRS and FWO (Belgium); CNPq, CAPES, FAPERJ, FAPERGS, and FAPESP (Brazil); MES and BNSF (Bulgaria); CERN; CAS, MoST, and NSFC (China); MINCIENCIAS (Colombia); MSES and CSF (Croatia); RIF (Cyprus); SENESCYT (Ecuador); MoER, ERC PUT and ERDF (Estonia); Academy of Finland, MEC, and HIP (Finland); CEA and CNRS/IN2P3 (France); BMBF, DFG, and HGF (Germany); GSRI (Greece); NKFIH (Hungary); DAE and DST (India); IPM (Iran); SFI (Ireland); INFN (Italy); MSIP and NRF (Republic of Korea); MES (Latvia); LAS (Lithuania); MOE and UM (Malaysia); BUAP, CINVESTAV, CONACYT, LNS, SEP, and UASLP-FAI (Mexico); MOS (Montenegro); MBIE (New Zealand); PAEC (Pakistan); MES and NSC (Poland); FCT (Portugal); MESTD (Serbia); MCIN/AEI and PCTI (Spain); MOSTR (Sri Lanka); Swiss Funding Agencies (Switzerland); MST (Taipei); MHESI and NSTDA (Thailand); TUBITAK and TENMAK (Turkey); NASU (Ukraine); STFC (United Kingdom); DOE and NSF (USA).
    
\hyphenation{Rachada-pisek} Individuals have received support from the Marie-Curie program and the European Research Council and Horizon 2020 Grant, contract Nos.\ 675440, 724704, 752730, 758316, 765710, 824093, 884104, and COST Action CA16108 (European Union); the Leventis Foundation; the Alfred P.\ Sloan Foundation; the Alexander von Humboldt Foundation; the Science Committee, project no. 22rl-037 (Armenia); the Belgian Federal Science Policy Office; the Fonds pour la Formation \`a la Recherche dans l'Industrie et dans l'Agriculture (FRIA-Belgium); the Agentschap voor Innovatie door Wetenschap en Technologie (IWT-Belgium); the F.R.S.-FNRS and FWO (Belgium) under the ``Excellence of Science -- EOS" -- be.h project n.\ 30820817; the Beijing Municipal Science \& Technology Commission, No. Z191100007219010; the Ministry of Education, Youth and Sports (MEYS) of the Czech Republic; the Shota Rustaveli National Science Foundation, grant FR-22-985 (Georgia); the Deutsche Forschungsgemeinschaft (DFG), under Germany's Excellence Strategy -- EXC 2121 ``Quantum Universe" -- 390833306, and under project number 400140256 - GRK2497; the Hellenic Foundation for Research and Innovation (HFRI), Project Number 2288 (Greece); the Hungarian Academy of Sciences, the New National Excellence Program - \'UNKP, the NKFIH research grants K 124845, K 124850, K 128713, K 128786, K 129058, K 131991, K 133046, K 138136, K 143460, K 143477, 2020-2.2.1-ED-2021-00181, and TKP2021-NKTA-64 (Hungary); the Council of Science and Industrial Research, India; the Latvian Council of Science; the Ministry of Education and Science, project no. 2022/WK/14, and the National Science Center, contracts Opus 2021/41/B/ST2/01369 and 2021/43/B/ST2/01552 (Poland); the Funda\c{c}\~ao para a Ci\^encia e a Tecnologia, grant CEECIND/01334/2018 (Portugal); the National Priorities Research Program by Qatar National Research Fund; MCIN/AEI/10.13039/501100011033, ERDF ``a way of making Europe", and the Programa Estatal de Fomento de la Investigaci{\'o}n Cient{\'i}fica y T{\'e}cnica de Excelencia Mar\'{\i}a de Maeztu, grant MDM-2017-0765 and Programa Severo Ochoa del Principado de Asturias (Spain); the Chulalongkorn Academic into Its 2nd Century Project Advancement Project, and the National Science, Research and Innovation Fund via the Program Management Unit for Human Resources \& Institutional Development, Research and Innovation, grant B05F650021 (Thailand); the Kavli Foundation; the Nvidia Corporation; the SuperMicro Corporation; the Welch Foundation, contract C-1845; and the Weston Havens Foundation (USA).
\end{acknowledgments}
\bibliography{auto_generated}
\clearpage
\appendix 
\section{Cross-experimental comparisons\label{app:Figures}}
\begin{figure}[!htb]\centering
    \includegraphics[width=0.6\textwidth]{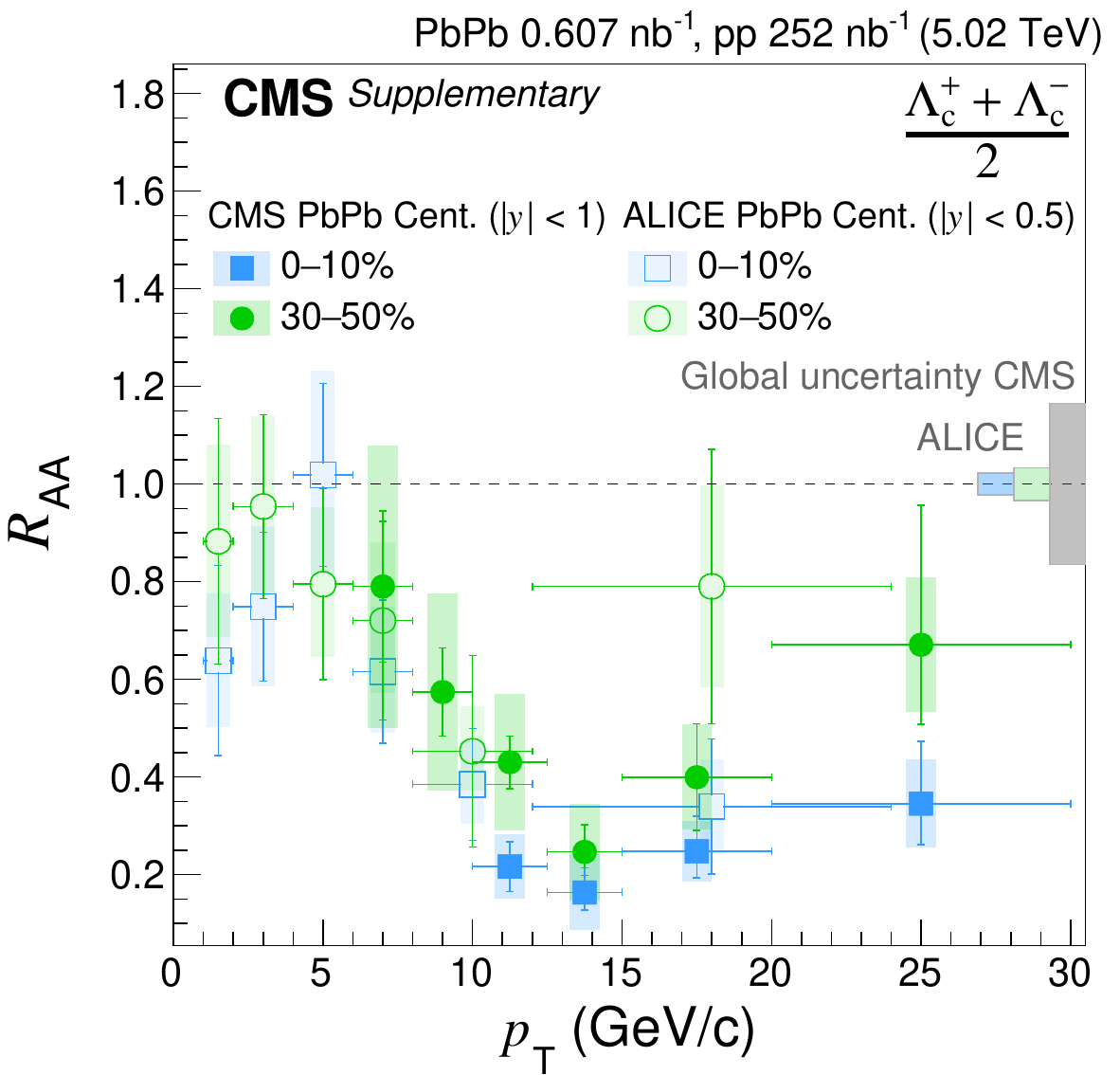}
    \caption{The nuclear modification factor $\RAA$ vs. \pt for \PcgLp production in \PbPb collisions. Results are shown for the 0--10 (closed squares) and 30--50\% (closed circles) centrality ranges. Also shown are the published results from the ALICE Collaboration~\cite{ALICE_PbPb2} for the same centrality ranges with open markers. The boxes and error bars represent the statistical and point-to-point systematic uncertainties, respectively. Global systematic uncertainties are indicated by the bands at unity. For CMS (gray band) this includes the uncertainties in the $\Pp\Pp$ collision luminosity, the number of \PbPb MB events, and the tracking efficiency. The global systematic uncertainties for the ALICE results (color bands) are shown separately for the two centrality ranges.}
    \label{app:Raa_plot_ALICE_comparison}
\end{figure}

\begin{figure}[!htb]\centering
    \includegraphics[width=0.6\textwidth]{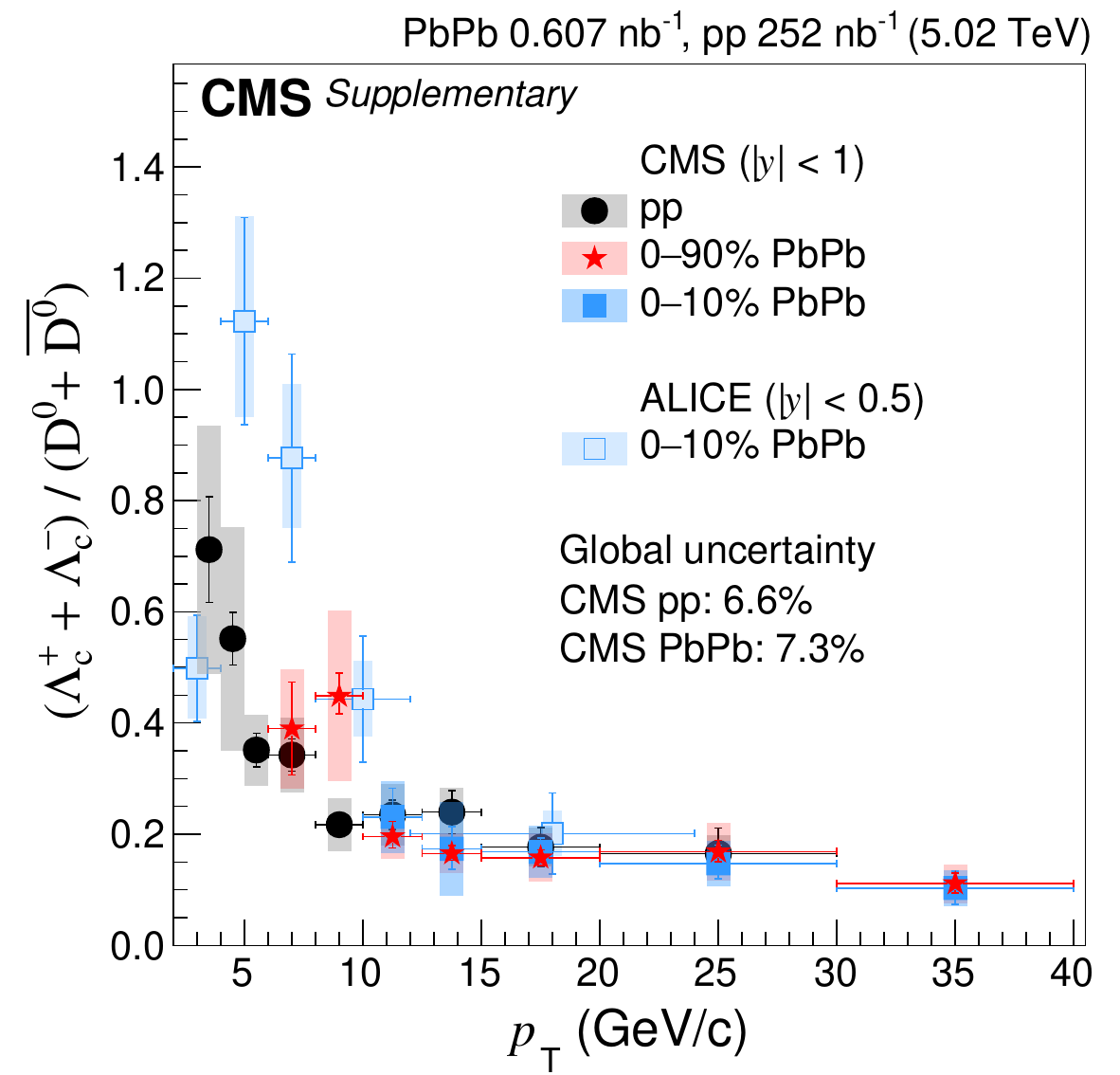}
    \caption{The ratio of the production cross sections of prompt $\PcgLp$ to prompt $\PDz$ versus \pt in $\Pp\Pp$ collisions is represented by the closed circles. The ratio for the 0--90 (closed stars) and 0--10\% (closed squares) centrality classes in \PbPb collisions are compared to the $\Pp\Pp$ result. The boxes and error bars represent the systematic and statistical uncertainties, respectively. The horizontal error bars represent the bin widths. The 6.6 and 7.3\% normalization uncertainties in $\Pp\Pp$ and \PbPb collisions, respectively, are not included in the boxes representing the systematic uncertainties for each data point. Results from the ALICE Collaboration~\cite{ALICE_PbPb2} for \PbPb collisions in the 0--10\% centrality class are also shown using the open square markers.}
    \label{app:LcD0_plot_ALICE_comparison}
\end{figure}
\cleardoublepage \section{The CMS Collaboration \label{app:collab}}\begin{sloppypar}\hyphenpenalty=5000\widowpenalty=500\clubpenalty=5000\input{HIN-21-004-public-authorlist.tex}\end{sloppypar}
\end{document}

%% file: HIN-21-004-public-authorlist.tex
\cmsinstitute{Yerevan Physics Institute, Yerevan, Armenia}
{\tolerance=6000
A.~Tumasyan\cmsAuthorMark{1}\cmsorcid{0009-0000-0684-6742}
\par}
\cmsinstitute{Institut f\"{u}r Hochenergiephysik, Vienna, Austria}
{\tolerance=6000
W.~Adam\cmsorcid{0000-0001-9099-4341}, J.W.~Andrejkovic, T.~Bergauer\cmsorcid{0000-0002-5786-0293}, S.~Chatterjee\cmsorcid{0000-0003-2660-0349}, K.~Damanakis\cmsorcid{0000-0001-5389-2872}, M.~Dragicevic\cmsorcid{0000-0003-1967-6783}, A.~Escalante~Del~Valle\cmsorcid{0000-0002-9702-6359}, P.S.~Hussain\cmsorcid{0000-0002-4825-5278}, M.~Jeitler\cmsAuthorMark{2}\cmsorcid{0000-0002-5141-9560}, N.~Krammer\cmsorcid{0000-0002-0548-0985}, L.~Lechner\cmsorcid{0000-0002-3065-1141}, D.~Liko\cmsorcid{0000-0002-3380-473X}, I.~Mikulec\cmsorcid{0000-0003-0385-2746}, P.~Paulitsch, J.~Schieck\cmsAuthorMark{2}\cmsorcid{0000-0002-1058-8093}, R.~Sch\"{o}fbeck\cmsorcid{0000-0002-2332-8784}, D.~Schwarz\cmsorcid{0000-0002-3821-7331}, M.~Sonawane\cmsorcid{0000-0003-0510-7010}, S.~Templ\cmsorcid{0000-0003-3137-5692}, W.~Waltenberger\cmsorcid{0000-0002-6215-7228}, C.-E.~Wulz\cmsAuthorMark{2}\cmsorcid{0000-0001-9226-5812}
% ...
\par}
\vskip\cmsinstskip
\dag:~Deceased\\
$^{1}$Also at Yerevan State University, Yerevan, Armenia\\
$^{2}$Also at TU Wien, Vienna, Austria\\
% ...


%% file: HIN-21-004_temp.bbl
\providecommand{\href}[2]{#2}\begingroup\raggedright\begin{thebibliography}{10}%
\makeatletter
\providecommand{\hrefCMSnoop }[0]{\@secondoftwo}%
\makeatother
\providecommand{\doi}{\texttt{doi:}\begingroup \urlstyle{tt}\Url}

\bibitem{QGP}
\href {http://jetp.ras.ru/cgi-bin/e/index/e/47/2/p212?a=list}{E.~V. Shuryak,
  ``Theory of hadronic plasma'',} \textit{ Sov. Phys. JETP} \textbf{ 47} (1978)
212.

\bibitem{Andronic:2015wma}
\hrefCMSnoop {}{A.~Andronic {et~al.}, ``{Heavy-flavour and quarkonium
  production in the LHC era: From proton\textendash{}proton to heavy-ion
  collisions}'',} \textit{ Eur. Phys. J. C} \textbf{ 76} (2016) 107,
  \href{http://dx.doi.org/10.1140/epjc/s10052-015-3819-5}{\doi{10.1140/epjc/s10052-015-3819-5}},
  \href{http://www.arXiv.org/abs/1506.03981}{\texttt{arXiv:1506.03981}}.

\bibitem{Prino:2016cni}
\hrefCMSnoop {}{F.~Prino and R.~Rapp, ``Open heavy flavor in {QCD} matter and
  in nuclear collisions'',} \textit{ J. Phys. G} \textbf{ 43} (2016) 093002,
  \href{http://dx.doi.org/10.1088/0954-3899/43/9/093002}{\doi{10.1088/0954-3899/43/9/093002}},
  \href{http://www.arXiv.org/abs/1603.00529}{\texttt{arXiv:1603.00529}}.

\bibitem{Dong:2019unq}
\hrefCMSnoop {}{X.~Dong and V.~Greco, ``{Heavy quark production and properties
  of Quark\textendash{}Gluon Plasma}'',} \textit{ Prog. Part. Nucl. Phys.}
  \textbf{ 104} (2019) 97,
  \href{http://dx.doi.org/10.1016/j.ppnp.2018.08.001}{\doi{10.1016/j.ppnp.2018.08.001}}.

\bibitem{Dong:2019byy}
\hrefCMSnoop {}{X.~Dong, Y.-J. Lee, and R.~Rapp, ``Open heavy-flavor production
  in heavy-ion collisions'',} \textit{ Ann. Rev. Nucl. Part. Sci.} \textbf{ 69}
  (2019) 417,
  \href{http://dx.doi.org/10.1146/annurev-nucl-101918-023806}{\doi{10.1146/annurev-nucl-101918-023806}},
  \href{http://www.arXiv.org/abs/1903.07709}{\texttt{arXiv:1903.07709}}.

\bibitem{Busza:2018rrf}
\hrefCMSnoop {}{W.~Busza, K.~Rajagopal, and W.~van~der Schee, ``Heavy ion
  collisions: The big picture, and the big questions'',} \textit{ Ann. Rev.
  Nucl. Part. Sci.} \textbf{ 68} (2018) 339,
  \href{http://dx.doi.org/10.1146/annurev-nucl-101917-020852}{\doi{10.1146/annurev-nucl-101917-020852}},
  \href{http://www.arXiv.org/abs/1802.04801}{\texttt{arXiv:1802.04801}}.

\bibitem{Apolin_rio_2022}
\hrefCMSnoop {}{L.~Apolin{\'{a}}rio, Y.-J. Lee, and M.~Winn, ``Heavy quarks and
  jets as probes of the {QGP}'',} \textit{ Prog. Part. Nucl. Phys.} \textbf{
  127} (2022) 103990,
  \href{http://dx.doi.org/10.1016/j.ppnp.2022.103990}{\doi{10.1016/j.ppnp.2022.103990}},
  \href{http://www.arXiv.org/abs/2203.16352}{\texttt{arXiv:2203.16352}}.

\bibitem{energyloss}
\hrefCMSnoop {}{A.~Beraudo {et~al.}, ``Extraction of heavy-flavor transport
  coefficients in {QCD} matter'',} \textit{ Nucl. Phys. A} \textbf{ 979} (2018)
  21,
  \href{http://dx.doi.org/10.1016/j.nuclphysa.2018.09.002}{\doi{10.1016/j.nuclphysa.2018.09.002}},
\href{http://www.arXiv.org/abs/1803.03824}{\texttt{arXiv:1803.03824}}.

\bibitem{ANDERSSON198331}
\hrefCMSnoop {}{B.~Andersson, G.~Gustafson, G.~Ingelman, and T.~Sj{\"o}strand,
  ``Parton fragmentation and string dynamics'',} \textit{ Phys. Rept.} \textbf{
  97} (1983) 31,
  \href{http://dx.doi.org/10.1016/0370-1573(83)90080-7}{\doi{10.1016/0370-1573(83)90080-7}}.

\bibitem{Greco_introduction}
\hrefCMSnoop {}{V.~Greco, C.~M. Ko, and P.~L{\'e}vai, ``Parton coalescence and
  the antiproton/pion anomaly at {RHIC}'',} \textit{ Phys. Rev. Lett.} \textbf{
  90} (2003) 202302,
  \href{http://dx.doi.org/10.1103/PhysRevLett.90.202302}{\doi{10.1103/PhysRevLett.90.202302}},
\href{http://www.arXiv.org/abs/nucl-th/0301093}{\texttt{arXiv:nucl-th/0301093}}.

\bibitem{Greco_follow}
\hrefCMSnoop {}{R.~J. Fries, V.~Greco, and P.~Sorensen, ``Coalescence models
  for hadron formation from quark gluon plasma'',} \textit{ Ann. Rev. Nucl.
  Part. Sci.} \textbf{ 58} (2008) 177,
  \href{http://dx.doi.org/10.1146/annurev.nucl.58.110707.171134}{\doi{10.1146/annurev.nucl.58.110707.171134}},
\href{http://www.arXiv.org/abs/0807.4939}{\texttt{arXiv:0807.4939}}.

\bibitem{Oh09}
\hrefCMSnoop {}{Y.~Oh, C.~M. Ko, S.~H. Lee, and S.~Yasui, ``Ratios of heavy
  baryons to heavy mesons in relativistic nucleus-nucleus collisions'',}
  \textit{ Phys. Rev. C} \textbf{ 79} (2009) 044905,
  \href{http://dx.doi.org/10.1103/PhysRevC.79.044905}{\doi{10.1103/PhysRevC.79.044905}},
\href{http://www.arXiv.org/abs/0901.1382}{\texttt{arXiv:0901.1382}}.

\bibitem{Lee}
S.~H. Lee\hrefCMSnoop {}{ {et~al.}, ``{$\Lambda_\text{c}$} enhancement from
  strongly coupled quark-gluon plasma'',} \textit{ Phys. Rev. Lett.} \textbf{
  100} (2008) 222301,
  \href{http://dx.doi.org/10.1103/PhysRevLett.100.222301}{\doi{10.1103/PhysRevLett.100.222301}},
\href{http://www.arXiv.org/abs/0709.3637}{\texttt{arXiv:0709.3637}}.

\bibitem{ghosh}
S.~Ghosh\hrefCMSnoop {}{ {et~al.}, ``{Diffusion of {$\Lambda_\text{c}$} in hot
  hadronic medium and its impact on {$\Lambda_\text{c}/D$} ratio}'',} \textit{
  Phys. Rev. D} \textbf{ 90} (2014) 054018,
  \href{http://dx.doi.org/10.1103/PhysRevD.90.054018}{\doi{10.1103/PhysRevD.90.054018}},
\href{http://www.arXiv.org/abs/1407.5069}{\texttt{arXiv:1407.5069}}.

\bibitem{Greco_publication}
S.~Plumari\hrefCMSnoop {}{ {et~al.}, ``Charmed hadrons from coalescence plus
  fragmentation in relativistic nucleus-nucleus collisions at {RHIC} and
  {LHC}'',} \textit{ Eur. Phys. J. C} \textbf{ 78} (2018) 348,
  \href{http://dx.doi.org/10.1140/epjc/s10052-018-5828-7}{\doi{10.1140/epjc/s10052-018-5828-7}},
\href{http://www.arXiv.org/abs/1712.00730}{\texttt{arXiv:1712.00730}}.

\bibitem{ALICE_pp}
\hrefCMSnoop {}{{ALICE Collaboration}, ``{$\Lambda_\text{c}^+$} production in
  pp collisions at $\sqrt{s} = 7$ {TeV} and in {p-Pb} collisions at {\sqrtsNN}
  = 5.02 {\TeV}'',} \textit{ JHEP} \textbf{ 04} (2018) 108,
  \href{http://dx.doi.org/10.1007/JHEP04(2018)108}{\doi{10.1007/JHEP04(2018)108}},
\href{http://www.arXiv.org/abs/1712.09581}{\texttt{arXiv:1712.09581}}.

\bibitem{ALICE_pp2}
\hrefCMSnoop {}{{ALICE Collaboration}, ``{$\Lambda_\text{c}^+$} production and
  baryon-to-meson ratios in pp and {p-Pb} collisions at {\sqrtsNN} = 5.02
  {\TeV} at the {LHC}'',} \textit{ Phys. Rev. Lett.} \textbf{ 127} (2021)
  202301,
  \href{http://dx.doi.org/10.1103/PhysRevLett.127.202301}{\doi{10.1103/PhysRevLett.127.202301}},
\href{http://www.arXiv.org/abs/2011.06078}{\texttt{arXiv:2011.06078}}.

\bibitem{HIN18009}
\hrefCMSnoop {}{{CMS Collaboration}, ``Production of {$\Lambda_\mathrm{c}^+$}
  baryons in proton-proton and lead-lead collisions at {\sqrtsNN} = 5.02
  {\TeV}'',} \textit{ Phys. Lett. B} \textbf{ 803} (2020) 135328,
  \href{http://dx.doi.org/10.1016/j.physletb.2020.135328}{\doi{10.1016/j.physletb.2020.135328}},
  \href{http://www.arXiv.org/abs/1906.03322}{\texttt{arXiv:1906.03322}}.

\bibitem{LHCb_pp}
\hrefCMSnoop {}{{LHCb Collaboration}, ``{Prompt charm production in pp
  collisions at $\sqrt{s} = 7$ TeV}'',} \textit{ Nucl. Phys. B} \textbf{ 871}
  (2013) 1,
  \href{http://dx.doi.org/10.1016/j.nuclphysb.2013.02.010}{\doi{10.1016/j.nuclphysb.2013.02.010}},
\href{http://www.arXiv.org/abs/1302.2864}{\texttt{arXiv:1302.2864}}.

\bibitem{GM-VFNS}
\hrefCMSnoop {}{M.~A.~G. Aivazis, J.~C. Collins, F.~I. Olness, and W.-K. Tung,
  ``{Leptoproduction} of heavy quarks. {II}. {A} unified {QCD} formulation of
  charged and neutral current processes from fixed-target to collider
  energies'',} \textit{ Phys. Rev. D} \textbf{ 50} (1994) 3102,
  \href{http://dx.doi.org/10.1103/PhysRevD.50.3102}{\doi{10.1103/PhysRevD.50.3102}}.

\bibitem{gmvfns1}
M.~Benzke\hrefCMSnoop {}{ {et~al.}, ``Prompt neutrinos from atmospheric charm
  in the general-mass variable-flavor-number scheme'',} \textit{ JHEP} \textbf{
  12} (2017) 021,
  \href{http://dx.doi.org/10.1007/JHEP12(2017)021}{\doi{10.1007/JHEP12(2017)021}},
\href{http://www.arXiv.org/abs/1705.10386}{\texttt{arXiv:1705.10386}}.

\bibitem{CRmodes}
\hrefCMSnoop {}{J.~R. Christiansen and P.~Z. Skands, ``String formation beyond
  leading colour'',} \textit{ JHEP} \textbf{ 08} (2015) 003,
  \href{http://dx.doi.org/10.1007/JHEP08(2015)003}{\doi{10.1007/JHEP08(2015)003}},
\href{http://www.arXiv.org/abs/1505.01681}{\texttt{arXiv:1505.01681}}.

\bibitem{LHCb_result}
\hrefCMSnoop {}{{LHCb Collaboration}, ``Prompt {$\Lambda^+_c$} production in
  {pPb} collisions at {\sqrtsNN} = 5.02 {\TeV}'',} \textit{ JHEP} \textbf{ 02}
  (2019) 102,
  \href{http://dx.doi.org/10.1007/JHEP02(2019)102}{\doi{10.1007/JHEP02(2019)102}},
\href{http://www.arXiv.org/abs/1809.01404}{\texttt{arXiv:1809.01404}}.

\bibitem{STAR_Lc}
\hrefCMSnoop {}{{STAR} Collaboration, ``First measurement of
  {$\Lambda_\text{c}^+$} baryon production in {Au+Au} collisions at {\sqrtsNN}
  = 200 {GeV}'',} \textit{ Phys. Rev. Lett.} \textbf{ 124} (2020) 172301,
  \href{http://dx.doi.org/10.1103/PhysRevLett.124.172301}{\doi{10.1103/PhysRevLett.124.172301}},
\href{http://www.arXiv.org/abs/1910.14628}{\texttt{arXiv:1910.14628}}.

\bibitem{PYTHIA8}
T.~Sj{\"o}strand\hrefCMSnoop {}{ {et~al.}, ``An introduction to {PYTHIA}
  8.2'',} \textit{ Comput. Phys. Commun.} \textbf{ 191} (2015) 159,
  \href{http://dx.doi.org/10.1016/j.cpc.2015.01.024}{\doi{10.1016/j.cpc.2015.01.024}},
\href{http://www.arXiv.org/abs/1410.3012}{\texttt{arXiv:1410.3012}}.

\bibitem{ALICE_PbPb}
\hrefCMSnoop {}{{ALICE Collaboration}, ``{$\Lambda_\text{c}^+$} production in
  {Pb-Pb} collisions at {\sqrtsNN} = 5.02 {\TeV}'',} \textit{ Phys. Lett. B}
  \textbf{ 793} (2019) 212,
  \href{http://dx.doi.org/10.1016/j.physletb.2019.04.046}{\doi{10.1016/j.physletb.2019.04.046}},
\href{http://www.arXiv.org/abs/1809.10922}{\texttt{arXiv:1809.10922}}.

\bibitem{ALICE_PbPb2}
\hrefCMSnoop {}{{ALICE Collaboration}, ``Constraining hadronization mechanisms
  with {$\Lambda_\text{c}^{+}/D^0$} production ratios in {Pb-Pb} collisions at
  {\sqrtsNN} = 5.02 {\TeV}'',} \textit{ Phys. Lett. B} \textbf{ 839} (2023)
  137796,
  \href{http://dx.doi.org/10.1016/j.physletb.2023.137796}{\doi{10.1016/j.physletb.2023.137796}},
\href{http://www.arXiv.org/abs/2112.08156}{\texttt{arXiv:2112.08156}}.

\bibitem{CMSX3872}
\hrefCMSnoop {}{{CMS Collaboration}, ``Evidence for {X(3872)} in {Pb-Pb}
  collisions and studies of its prompt production at {\sqrtsNN} = 5.02
  {\TeV}'',} \textit{ Phys. Rev. Lett.} \textbf{ 128} (2022) 032001,
  \href{http://dx.doi.org/10.1103/PhysRevLett.128.032001}{\doi{10.1103/PhysRevLett.128.032001}},
  \href{http://www.arXiv.org/abs/2102.13048}{\texttt{arXiv:2102.13048}}.

\bibitem{HIN20004}
\hrefCMSnoop {}{{CMS Collaboration}, ``Observation of the ${B}_\mathrm{c}^{+}$
  meson in {Pb-Pb} and {pp} collisions at {\sqrtsNN} = 5.02 {\TeV} and
  measurement of its nuclear modification factor'',} \textit{ Phys. Rev. Lett.}
  \textbf{ 128} (2022) 252301,
  \href{http://dx.doi.org/10.1103/PhysRevLett.128.252301}{\doi{10.1103/PhysRevLett.128.252301}},
  \href{http://www.arXiv.org/abs/2201.02659}{\texttt{arXiv:2201.02659}}.

\bibitem{CMS:2022bjp}
\href {http://cds.cern.ch/record/2809613}{{CMS Collaboration}, ``{CMS}
  luminosity measurement using nucleus-nucleus collisions at {\sqrtsNN} = 5.02
  {\TeV} in 2018'',} CMS Physics Analysis Summary CMS-PAS-LUM-18-001, 2022.

\bibitem{CMS-PAS-LUM-19-001}
\href {http://cds.cern.ch/record/2765655}{{CMS Collaboration}, ``Luminosity
  measurement in proton-proton collisions at 5.02 {TeV} in 2017 at {CMS}'',}
  CMS Physics Analysis Summary CMS-PAS-LUM-19-001, 2021.

\bibitem{Sirunyan:2759951}
\hrefCMSnoop {}{{CMS Collaboration}, ``Precision luminosity measurement in
  proton-proton collisions at $\sqrt{s} =$ 13 {TeV} in 2015 and 2016 at
  {CMS}'',} \textit{ Eur. Phys. J. C} \textbf{ 81} (2021) 800,
  \href{http://dx.doi.org/10.1140/epjc/s10052-021-09538-2}{\doi{10.1140/epjc/s10052-021-09538-2}},
  \href{http://www.arXiv.org/abs/2104.01927}{\texttt{arXiv:2104.01927}}.

\bibitem{Rapp_ppresult}
\hrefCMSnoop {}{M.~He and R.~Rapp, ``Charm-baryon production in proton-proton
  collisions'',} \textit{ Phys. Lett. B} \textbf{ 795} (2019) 117,
  \href{http://dx.doi.org/10.1016/j.physletb.2019.06.004}{\doi{10.1016/j.physletb.2019.06.004}},
\href{http://www.arXiv.org/abs/1902.08889}{\texttt{arXiv:1902.08889}}.

\bibitem{HIN16001}
\hrefCMSnoop {}{{CMS Collaboration}, ``Nuclear modification factor of {D$^0$}
  mesons in {PbPb} collisions at {\sqrtsNN} = 5.02 {\TeV}'',} \textit{ Phys.
  Lett. B} \textbf{ 782} (2018) 474,
  \href{http://dx.doi.org/10.1016/j.physletb.2018.05.074}{\doi{10.1016/j.physletb.2018.05.074}},
\href{http://www.arXiv.org/abs/1708.04962}{\texttt{arXiv:1708.04962}}.

\bibitem{hepdata}
\hrefCMSnoop {}{}{HEPD}ata record for this analysis, 2023.
\newblock
  \href{http://dx.doi.org/10.17182/hepdata.135973}{\doi{10.17182/hepdata.135973}}.

\bibitem{ppreconstruct}
\hrefCMSnoop {}{{CMS Collaboration}, ``Description and performance of track and
  primary-vertex reconstruction with the {CMS} tracker'',} \textit{ JINST}
  \textbf{ 9} (2014) P10009,
  \href{http://dx.doi.org/10.1088/1748-0221/9/10/P10009}{\doi{10.1088/1748-0221/9/10/P10009}},
\href{http://www.arXiv.org/abs/1405.6569}{\texttt{arXiv:1405.6569}}.

\bibitem{CMS:detector}
\hrefCMSnoop {}{{CMS Collaboration}, ``The {CMS} experiment at the {CERN}
  {LHC}'',} \textit{ JINST} \textbf{ 3} (2008) S08004,
\href{http://dx.doi.org/10.1088/1748-0221/3/08/S08004}{\doi{10.1088/1748-0221/3/08/S08004}}.

\bibitem{PbPbreconstruct}
\hrefCMSnoop {}{{CMS Collaboration}, ``Charged-particle nuclear modification
  factors in {PbPb} and {pPb} collisions at {\sqrtsNN} = 5.02 {\TeV}'',}
  \textit{ JHEP} \textbf{ 04} (2017) 039,
  \href{http://dx.doi.org/10.1007/JHEP04(2017)039}{\doi{10.1007/JHEP04(2017)039}},
\href{http://www.arXiv.org/abs/1611.01664}{\texttt{arXiv:1611.01664}}.

\bibitem{ptdistribution}
\hrefCMSnoop {}{{CMS Collaboration}, ``Transverse-momentum and pseudorapidity
  distributions of charged hadrons in pp collisions at $\sqrt{s} = 0.9$ and
  2.36 {\TeV}'',} \textit{ JHEP} \textbf{ 02} (2010) 041,
  \href{http://dx.doi.org/10.1007/JHEP02(2010)041}{\doi{10.1007/JHEP02(2010)041}},
\href{http://www.arXiv.org/abs/1002.0621}{\texttt{arXiv:1002.0621}}.

\bibitem{CMS:2016ngn}
\hrefCMSnoop {}{{CMS Collaboration}, ``{The CMS trigger system}'',} \textit{
  JINST} \textbf{ 12} (2017) P01020,
  \href{http://dx.doi.org/10.1088/1748-0221/12/01/P01020}{\doi{10.1088/1748-0221/12/01/P01020}},
  \href{http://www.arXiv.org/abs/1609.02366}{\texttt{arXiv:1609.02366}}.

\bibitem{Eventune}
\hrefCMSnoop {}{{CMS Collaboration}, ``Event generator tunes obtained from
  underlying event and multiparton scattering measurements'',} \textit{ Eur.
  Phys. J. C} \textbf{ 76} (2016) 155,
  \href{http://dx.doi.org/10.1140/epjc/s10052-016-3988-x}{\doi{10.1140/epjc/s10052-016-3988-x}},
\href{http://www.arXiv.org/abs/1512.00815}{\texttt{arXiv:1512.00815}}.

\bibitem{jetquench}
\hrefCMSnoop {}{I.~P. Lokhtin and A.~M. Snigirev, ``A model of jet quenching in
  ultrarelativistic heavy ion collisions and high-\pt hadron spectra at
  {RHIC}'',} \textit{ Eur. Phys. J. C} \textbf{ 45} (2006) 211,
  \href{http://dx.doi.org/10.1140/epjc/s2005-02426-3}{\doi{10.1140/epjc/s2005-02426-3}},
\href{http://www.arXiv.org/abs/hep-ph/0506189}{\texttt{arXiv:hep-ph/0506189}}.

\bibitem{Eventgen}
\hrefCMSnoop {}{D.~J. Lange, ``The {EvtGen} particle decay simulation
  package'',} \textit{ Nucl. Instrum. Meth. A} \textbf{ 462} (2001) 152,
\href{http://dx.doi.org/10.1016/S0168-9002(01)00089-4}{\doi{10.1016/S0168-9002(01)00089-4}}.

\bibitem{Workman:2022ynf}
\hrefCMSnoop {}{{Particle Data Group} Collaboration, ``Review of particle
  physics'',} \textit{ PTEP} \textbf{ 2022} (2022) 083C01,
  \href{http://dx.doi.org/10.1093/ptep/ptac097}{\doi{10.1093/ptep/ptac097}}.

\bibitem{GEANT4}
\hrefCMSnoop {}{{GEANT4} Collaboration, ``{\GEANTfour}---a simulation
  toolkit'',} \textit{ Nucl. Instrum. Meth. A} \textbf{ 506} (2003) 250,
\href{http://dx.doi.org/10.1016/S0168-9002(03)01368-8}{\doi{10.1016/S0168-9002(03)01368-8}}.

\bibitem{LLR}
\hrefCMSnoop {}{E.~L. Lehmann, ``On likelihood ratio tests'',} in \textit{
  Institute of Mathematical Statistics Lecture Notes - Monograph Series},
  volume~49, p.~1.
\newblock Institute of Mathematical Statistics, 2006.
\newblock
  \href{http://www.arXiv.org/abs/math/0610835}{\texttt{arXiv:math/0610835}}.
\newblock
  \href{http://dx.doi.org/10.1214/074921706000000356}{\doi{10.1214/074921706000000356}}.

\bibitem{FONLL}
\hrefCMSnoop {}{M.~Cacciari, M.~Greco, and P.~Nason, ``The \pt spectrum in
  heavy flavor hadroproduction'',} \textit{ JHEP} \textbf{ 05} (1998) 007,
  \href{http://dx.doi.org/10.1088/1126-6708/1998/05/007}{\doi{10.1088/1126-6708/1998/05/007}},
\href{http://www.arXiv.org/abs/hep-ph/9803400}{\texttt{arXiv:hep-ph/9803400}}.

\bibitem{bhadron_data_comp}
\hrefCMSnoop {}{{CMS Collaboration}, ``Measurement of the differential
  inclusive $\mathrm{B}^{+}$ hadron cross sections in pp collisions at
  $\sqrt{s}=13\text{ } \mathrm {TeV}$'',} \textit{ Phys. Lett. B} \textbf{ 771}
  (2017) 435,
  \href{http://dx.doi.org/10.1016/j.physletb.2017.05.074}{\doi{10.1016/j.physletb.2017.05.074}},
  \href{http://www.arXiv.org/abs/1609.00873}{\texttt{arXiv:1609.00873}}.

\bibitem{Glauber}
\hrefCMSnoop {}{M.~L. Miller, K.~Reygers, S.~J. Sanders, and P.~Steinberg,
  ``Glauber modeling in high-energy nuclear collisions'',} \textit{ Ann. Rev.
  Nucl. Part. Science} \textbf{ 57} (2007) 205,
  \href{http://dx.doi.org/10.1146/annurev.nucl.57.090506.123020}{\doi{10.1146/annurev.nucl.57.090506.123020}},
  \href{http://www.arXiv.org/abs/nucl-ex/0701025}{\texttt{arXiv:nucl-ex/0701025}}.

\bibitem{HIN16011}
\hrefCMSnoop {}{{CMS Collaboration}, ``Measurement of the {${B}^{\pm}$} meson
  nuclear modification factor in {Pb-Pb} collisions at {\sqrtsNN} = 5.02
  {\TeV}'',} \textit{ Phys. Rev. Lett.} \textbf{ 119} (2017) 152301,
  \href{http://dx.doi.org/10.1103/PhysRevLett.119.152301}{\doi{10.1103/PhysRevLett.119.152301}},
\href{http://www.arXiv.org/abs/1705.04727}{\texttt{arXiv:1705.04727}}.

\bibitem{gmvfns2}
\hrefCMSnoop {}{B.~Kniehl, G.~Kramer, I.~Schienbein, and H.~Spiesberger,
  ``{\PcgLp} production in pp collisions with a new fragmentation function'',}
  \textit{ Phys. Rev. D} \textbf{ 101} (2020) 114021,
  \href{http://dx.doi.org/10.1103/PhysRevD.101.114021}{\doi{10.1103/PhysRevD.101.114021}},
\href{http://www.arXiv.org/abs/2004.04213}{\texttt{arXiv:2004.04213}}.

\bibitem{OPAL:LC_Study}
\hrefCMSnoop {}{{OPAL} Collaboration, ``A study of charm hadron production in
  {$Z^0 \to c\bar c$} and {$Z^0 \to b\bar b$} decays at {LEP}'',} \textit{ Z.
  Phys. C - Particles and Fields} \textbf{ 72} (1996) 1,
  \href{http://dx.doi.org/10.1007/BF02909127}{\doi{10.1007/BF02909127}}.

\bibitem{Belle:LC_Study}
\hrefCMSnoop {}{{Belle} Collaboration, ``Charm hadrons from fragmentation and
  $b$ decays in ${e}^{+}{e}^{\ensuremath{-}}$ annihilation at
  $\sqrt{s}=10.6\text{ }\text{ }\mathrm{GeV}$'',} \textit{ Phys. Rev. D}
  \textbf{ 73} (2006) 032002,
  \href{http://dx.doi.org/10.1103/PhysRevD.73.032002}{\doi{10.1103/PhysRevD.73.032002}},
  \href{http://www.arXiv.org/abs/hep-ex/0506068}{\texttt{arXiv:hep-ex/0506068}}.

\bibitem{HIN16016}
\hrefCMSnoop {}{{CMS Collaboration}, ``Studies of beauty suppression via
  nonprompt {\PDz} mesons in {Pb-Pb} collisions at {\sqrtsNN} = 5.02 {\TeV}'',}
  \textit{ Phys. Rev. Lett.} \textbf{ 123} (2019) 022001,
  \href{http://dx.doi.org/10.1103/PhysRevLett.123.022001}{\doi{10.1103/PhysRevLett.123.022001}},
\href{http://www.arXiv.org/abs/1810.11102}{\texttt{arXiv:1810.11102}}.

\bibitem{Greco_pub_2}
\hrefCMSnoop {}{V.~Minissale, S.~Plumari, and V.~Greco, ``Charm hadrons in pp
  collisions at {LHC} energy within a coalescence plus fragmentation
  approach'',} \textit{ Phys. Lett. B} \textbf{ 821} (2021) 136622,
  \href{http://dx.doi.org/10.1016/j.physletb.2021.136622}{\doi{10.1016/j.physletb.2021.136622}},
\href{http://www.arXiv.org/abs/2012.12001v3}{\texttt{arXiv:2012.12001v3}}.

\bibitem{Rapp_PbPb}
\hrefCMSnoop {}{M.~He and R.~Rapp, ``Hadronization and charm-hadron ratios in
  heavy-ion collisions'',} \textit{ Phys. Rev. Lett.} \textbf{ 124} (2020)
  042301,
  \href{http://dx.doi.org/10.1103/PhysRevLett.124.042301}{\doi{10.1103/PhysRevLett.124.042301}},
\href{http://www.arXiv.org/abs/1905.09216v2}{\texttt{arXiv:1905.09216v2}}.

\end{thebibliography}\endgroup
